\documentclass[10pt,a4paper,oneside]{article}
\pdfoutput=1
\usepackage[english]{babel}
\usepackage{longtable}

\usepackage{rotating,longtable}

\usepackage{amsmath,amsfonts,amssymb,bm,mathrsfs}\interdisplaylinepenalty=2500

\usepackage{graphicx,color,rotating,epstopdf}
\makeatletter\def\sgn{\mathop{\operator@font sgn}\nolimits}\makeatother
\newcommand{\unit}[1]{\ensuremath{\mathrm{#1}}}

\newcommand{\req}[1]{(\ref{#1})}
\newcommand{\sidenote}[1]{\marginpar{\sf\bfseries\small\flushleft#1}}
\graphicspath{{figs/},{figs-old/}}			
\def\PrintGraphicFileName{1}			
\newcommand{\namedgraphics}[3]{\parbox{#3}{\ifnum\PrintGraphicFileName>0\rotatebox{90}{\smash{\ttfamily\scriptsize\raisebox{0.8em}{#2}}}\fi\hspace*{\fill}\includegraphics[#1]{#2}\hspace*{\fill}}}
\newcommand{\namedcombographics}[3]{\parbox{#3}{\ifnum
\PrintGraphicFileName>0\rotatebox{90}{\smash{\ttfamily\scriptsize\raisebox{1em}{#2}}}\fi
\hspace*{\fill}\scalebox{#1}{
\ifnum\pdfoutput=0\input{#2.pstex_t}\else\input{#2.pdftex_t}\fi}\hspace*{\fill}}}

\newcommand{\TodayHeader}{\today}
\newcommand{\TodayFront}{\today}
\title{The cross-spectrum experimental method}
\author{Enrico Rubiola$^\bigstar$ and Fran\c{c}ois Vernotte$^\blacklozenge$\\
\small web page \texttt{http://rubiola.org}
\\[4em]\includegraphics[width=0.35\textwidth]{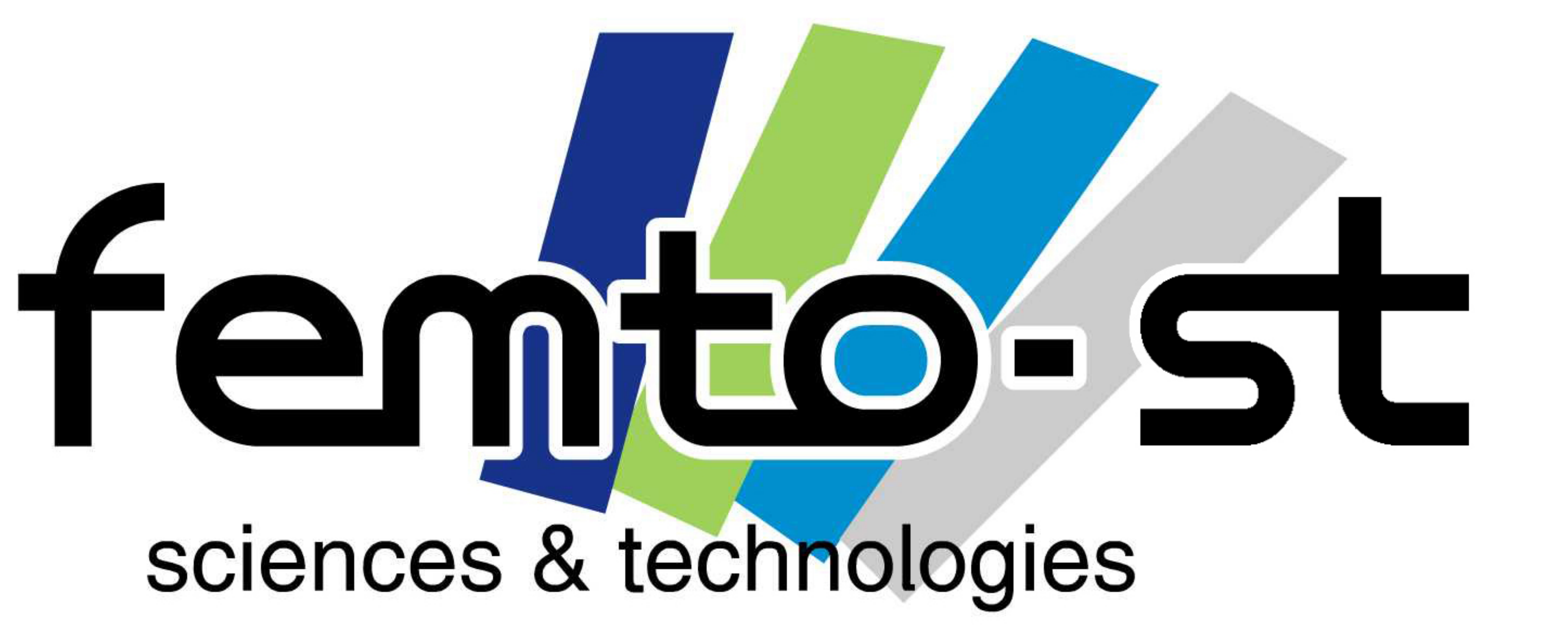}\\[0.5em]
\small $\bigstar$ CNRS/UFC FEMTO-ST Institute, Besan\c{c}on, France\\
\small $\blacklozenge$ CNRS/UFC UTINAM Besancon Observatory, France\\[1.5em]}
\date{\small\TodayFront}
\def\pageheads{E. Rubiola, The cross-spectrum experimental method \hfill~\TodayHeader\hfill}
\begin{document}
\maketitle

\begin{abstract}
The noise of a device under test (DUT) is measured simultaneously with two instruments, each of which contributes its own background.  The average cross power spectral density converges to the DUT power spectral density.  This method enables the extraction of the DUT noise spectrum, even if it is significantly lower than the background.  After a snapshot on practical experiments, we go through the statistical theory and the choice of the estimator.  A few experimental techniques are described, with reference to phase noise and amplitude noise in RF/microwave systems and in photonic systems.
The set of applications of this method is wide.  The final section gives a short panorama on 
radio-astronomy, radiometry, quantum optics, thermometry (fundamental and applied), semiconductor technology, metallurgy, etc.

This report is intended as a tutorial, as opposed to a report on advanced research, yet addressed to a broad readership: technicians, practitioners, Ph.D. students, academics, and full-time scientists.         
\end{abstract}

\clearpage
\tableofcontents
\vspace{6em}

\section*{Notation}
\addcontentsline{toc}{section}{Notation}
\begin{center}
\begin{longtable}[t]{ll}
\bfseries Symbol	& \bfseries Meaning\\\hline
$a(t)\leftrightarrow A(f)$ & background noise of the instrument A\\
$b(t)\leftrightarrow B(f)$ & background noise of the instrument B\\
$c(t)\leftrightarrow C(f)$ & DUT noise, i.e., the useful signal\\
$b_i$ & coefficients of the power-law approximation of $S_\varphi(f)$\\&(in AM-PM noise)\\
%
$\mathrm{dev}\{\,\}$	& deviation, $\mathrm{dev}\{x\}=\smash{\sqrt{\mathbb{V}\{x\}}}$\\
$\mathbb{E}\{\,\}$	& mathematical expectation\\
$f$		& Fourier frequency, Hz\\
$f(x)$	& probability density function (PDF)\\
$F(x)$	& cumulative density function (CDF)\\
$\mathcal{F}\{\,\}$	& Fourier thansform operator\\
$h_i$ & coefficients of the power-law model of $S_\alpha(f)$ or $S_y(f)$\\&(in AM-PM noise)\\
$i$	& integer number, often as as an index\\
$\imath$	& imaginary unit, $\imath^2=-1$\\
$\Im\{\,\}$	& imaginary part of a complex quantity, as in $X''=\Im\{X\}$\\
$m$ & number of averaged spectra, as in $\left<|S_{yx}|\right>_m$\\
$O(\,)$		& order of, as in $e^x=1+x+O(x^2)$\\
$\mathbb{P}\{\,\}$	& probability, as in $\mathbb{P}\{x>0\}$\\
$P_N$	& probability that a value is negative, as in $P_N=\mathbb{P}\{x<0\}$\\
$P_P$	& probability that a value is positive, as in $P_P=\mathbb{P}\{x>0\}$\\
$R_{xx}(t')$ & autocorrelation function\\
$\Re\{\,\}$	& real part of a complex quantity, as in $X'=\Re\{X\}$\\
$S_{xx}(f)$	& PSD of the quantity $x$\\ 
$S_{yx}(f)$	& cross PSD of the quantities $y$ and $x$\\ 
$t$		& time\\
$T$	& measurement time\\
$\mathbb{V}\{\,\}$	& variance, mathematical expectation of\\
$x(t)\leftrightarrow X(f)$ & generic variable\\
$x(t)\leftrightarrow X(f)$ & signal at the FFT analyzer input, channel 1\\
$\mathbf{x}(t)$, $\mathbf{y}(t)$ & stochastic processes, of which $x(t)$ and $x(t)$ are realizations\\
$y(t)\leftrightarrow Y(f)$ & generic variable\\
$y(t)\leftrightarrow Y(f)$ & signal at the FFT analyzer input, channel 2\\
$\alpha(t)\leftrightarrow\mathcal{A}(f)$ & normalized-amplitude noise (in AM-PM noise)\\
$\Gamma(x)$& the gamma function used in probability\\
$\kappa^2$	& PSD of the signal $c(t)$\\
$\mu$	& average (the value of)\\
$\nu$	& frequency (Hz), used for carrier signals (in AM-PM noise)\\
$\nu$	& no.\ of degrees of freedom, in probability functions\\
$\sigma(\tau)$& Allan deviation, $\sqrt{\text{Allan variance}}$ (in AM-PM noise)\\
$\tau$	& measurement time of the Allan variance (in AM-PM noise)\\
$\varphi(t)\leftrightarrow\Phi(f)$ & phase noise (in AM-PM noise)\\
$\chi^2$	& in probability, $\chi^2=\mathbf{x}_1^2+\mathbf{x}_2^2+\mathbf{x}_3^2+\ldots$\ originates the\\& $\chi^2$ distribution\\[1em]
%
\bfseries Subscript	& \bfseries Meaning\\\hline
$T$	& truncated over the meas.\ time $T$, as in $x_T(t)$, $X_T(f)$\\[1em]
\bfseries Superscript	& \bfseries Meaning\\\hline
$\ast$	& complex conjugate, as in $|X|^2=XX^\ast$\\[1em]
\bfseries Symbol	& \bfseries Meaning\\\hline
$\left<~\right>$	& average.  Also $\left<~\right>_m$ average of $m$ values\\
$\hat{~}$ & estimator of a quantity, as in  $\smash{\hat{S}}_{yx}=\left<S_{yx}\right>_m$ \\
$'$, $''$ & real and imaginary part, as in $X=X'+\imath X''$\\
$\leftrightarrow$& transform inverse-transform pair, as in $x(t)\leftrightarrow X(s)$\\
$\dot{~}$	& time-derivative, as in $\dot{\varphi}(t)$ (in AM-PM noise)\\[1em]
%
\bfseries Acronym	& \bfseries Meaning\\\hline
AM	& Amplitude Modulation, often `AM noise' (in AM-PM noise)\\
CDF	& Cumulative Density Function\\
DUT	& Device Under Test\\
FFT	& Fast Fourier Transform\\
PM	& Phase Modulation, often `PM noise' (in AM-PM noise)\\
PDF	& Probability Density Function\\
PLL	& Phase Locked Loop (in AM-PM noise)\\
PSD	& (single-side) Power Spectral Density\\[1em]
%
\bfseries font/case & \bfseries Meaning\\\hline
uppercase & Fourier transform of the lower-case function\\
rm-bf & stochastic processes, as in $x(t)$ is a realization of $\mathbf{x}(t)$\\\hline
\multicolumn{2}{l}{Font/case is used in this way only in some special (and obvious) cases}
\end{longtable}
\end{center}

\clearpage

\section{Introduction}\label{sec:xsp-introduction}
Measuring a device under test (DUT), the observed spectrum contains the DUT noise, which we can call \emph{signal} because it is the object of the measurement, and the background noise of the instrument.  The core of the cross-spectrum measurement method is that we can measure the DUT simultaneously with two equal instruments.  Provided that experimental skill and a pinch of good luck guarantee that DUT and instruments are statistically independent, statistics enables to extract the DUT spectrum from the background.  

\begin{figure}[b]
\centering\namedgraphics{scale=0.6}{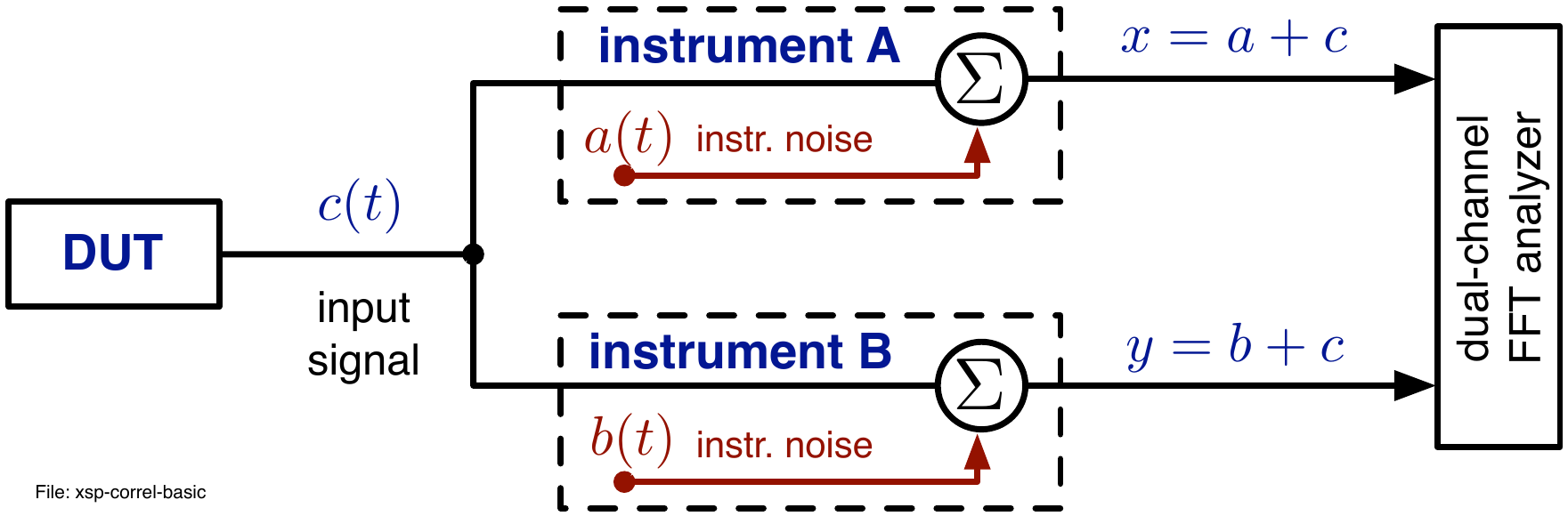}{\textwidth}
\caption{Basics of the cross-spectrum method.}
\label{fig:xsp-correl-basics}
\end{figure}
\begin{figure}
\centering\namedgraphics{scale=0.8}{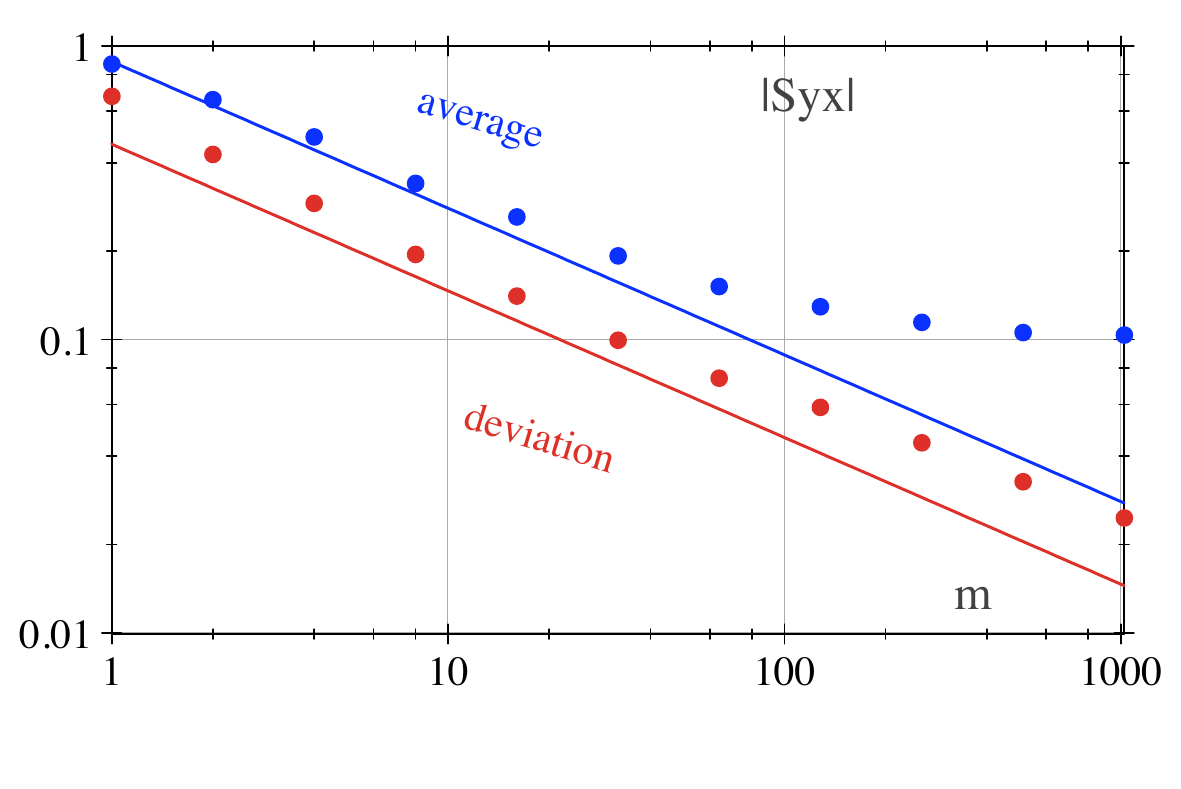}{\textwidth}
\vspace*{-2em}
\caption{Average and deviation of the cross spectrum $|\left<S_{yx}\right>_m|$, as a function of the number $m$ of averaged realizations of white Gaussian noise.  Since the statistical properties of $S_{yx}(f)$ are the same at any frequency, only one point (i.e., one frequency) is shown and the variable $f$ is dropped.
The DUT noise is 10 dB lower than the background.}
\label{fig:mce-sqrt-law}
\end{figure}
The two-channel measurement can be modeled as the block diagram of Fig.~\ref{fig:xsp-correl-basics}, where $a(t)$ and $b(t)$ are the background of the two instruments, and $c(t)$ the DUT noise, under the hypothesis that $a(t)$, $b(t)$ and $c(t)$ are statistically independent.  Thus, the observed signals are
\begin{align*}
x(t)&=c(t)+a(t)\\
y(t)&=c(t)+b(t)~.
\end{align*}
We are interested in the power spectral density\footnote{The PSD as a statistical concept will be defined afterwards.  Newcomers can provisionally use $S_{yx}(f)=\frac1TY(f)X^\ast(f)$, which is the is the readout of the FFT analyzer. $T$ is the measurement time.} (PSD), which is a normalized form of spectrum that expresses the power per unit of bandwidth, denoted with $S(f)$.
It will be shown that the average cross-PSD $\left<S_{yx}(f)\right>$ converges to the DUT PSD $S_{cc}(f)$, which is what we want to measure.  

The idea of the cross-spectrum method is explained in Fig.~\ref{fig:mce-sqrt-law}.  This figure builds from the output of the free-running analyzer, after selecting one frequency ($f_0$).  This is a sequence of $|S_{yx}(f_0)|$ called realizations, which we average on contiguous groups of $m$ values $|\left<S_{yx}(f)\right>_m|$.  The averages form a (slower) sequence whose statistical properties depend on $m$.  So, Fig.~\ref{fig:mce-sqrt-law} plots the average and the variance of the sequence of averages, as a function of $m$.
At small values of $m$, the background is dominant and decreases as $m$ increases.  Beyond $m\approx100$,  we observe that  $|\left<S_{yx}(f)\right>_m|$ stops decreasing and approaches the value of 0.1 ($-10$ dB), which is the DUT noise in this example.  The standard deviation further  decreases.  The background is dominant below $m\approx100$.  Beyond, the DUT noise shows up and the estimation accuracy increases, as seen from the deviation-to-average ratio.
Notice that the choice of $|\left<S_{yx}(f)\right>_m|$ as an estimator of $S_{yx}(f)$ is still arbitrary and will be further discussed.

All this report is about how and why the cross-spectrum converges to the DUT noise $S_{cc}(f)$, and about how this fact can be used in the laboratory practice.
The scheme of Fig.~\ref{fig:xsp-correl-basics} is analyzed from the following standpoints
\begin{description}
\item[Normal use.] All the noise processes [$a(t)$, $b(t)$ and $c(t)$] have non-negligible power.  We use the statistics to extract $S_{cc}(f)$.
\item[Statistical limit.]  In the absence of correlated phenomenon, thus with $c=0$, the average cross spectrum takes a finite nonzero value, limited by the number of averaged realizations. 
\item[Hardware limit.]  After removing the DUT, a (small) correlated part remain.  This phenomenon, due to crosstalk or to other effects, limits the instrument sensitivity.
\end{description}

Though the author is inclined to use phase and amplitude noise as the favorite examples (Section \ref{ssec:xsp-pm-noise} and \ref{ssec:xsp-am-noise}), the cross-spectrum method is of far more general interest.  Examples from a variety of research fields will be discussed in Section~\ref{ssec:xsp-other-applications}.

As a complement to this report, the reader is encouraged to refer to classical textbooks of probability and statistics, among which \cite{Feller:probability,Papoulis:probability,Cramer:statistics,Davenport-Root:noise} are preferred.

\section{Power spectral density}
The processes we describe are stationary and ergodic.  The requirement that noise be stationary and ergodic is not a stringent constraint in the laboratory practice because the words `stationary' and `ergodic' are the equivalent of `repeatable' and `reproducible' in experimental physics.  Thus, a realization $x(t)$ has the same statistical properties independently of the origin of time, and also the statistical properties of the entire process $\mathbf{x}(t)$.  Unless otherwise specified, $\mathbf{x}(t)$ is a zero-mean finite-power process.
The power spectral density (PSD) of such processes is 
\begin{align}
S_{xx}(f) &= \mathcal{F}\left\{R_{xx}(t')\right\}
\label{eqn:xsp-psd-def}
\end{align}
where $\mathcal{F}\{\:\}$ is the Fourier transform operator, 
\begin{align}
R_{xx}(t') &= \mathbb{E}\left\{ \mathbf{x}(t) \, \mathbf{x}(t+t')\right\}
\end{align}
the autocorrelation function, and $\mathbb{E}\{\:\}$ the mathematical expectation. 

As a simplified notation, we use the upper case for the Fourier transform, and the left-right arrow for the transform inverse-transform pair, thus 
\begin{align*}
x(t)\leftrightarrow X(f)\qquad\text{Fourier transform -- inverse transform pair}~.
\end{align*}
The two-sided Fourier transform and spectra are generally preferred in theoretical issues, while the experimentalist often prefers the single-sided representation.  Though we use the one-sided representation in all figures, often we do not need the distinction between one-sided and two-sided representation.  In most practical measurements the Fast Fourier Transform (FFT) replaces the traditional Fourier transform, and the frequency is a discrete variable.

The Wiener-Khintchine theorem for ergodic and stationary processes enables to calculate the PSD through the absolute value of the Fourier transform.  Thus it holds that 
\begin{align}
\label{eqn:xsp-psd-wk}
\mathbb{E}\left\{S_{xx}(f)\right\} 
&=\mathbb{E}\Bigl\{\lim_{T\rightarrow\infty}\Bigl[\frac1T\,X_T(f)\,X_T^\ast(f)\Bigr]\Bigr\}\\
&=\mathbb{E}\Bigl\{\lim_{T\rightarrow\infty}\Bigl[\frac1T\,\left|X_T(f)\right|^2\Bigr]\Bigr\}~,
\label{eqn:xsp-psd-wk-abs}
\end{align}
where the subscript $T$ means truncated over the measurement time $T$, and the superscript `$\ast$' stands for complex conjugate.  By the way, the factor $\frac1T$ is necessary for $S_{xx}(f)$ to have the physical dimension of a \emph{power density}, i.e., power per unit of frequency.

Omitting the expectation, \req{eqn:xsp-psd-wk} can be seen as a realization of the PSD\@.   In actual experiments the expectation is replaced with the average on a suitable number $m$ of spectrum samples
\begin{align}
&\left<S_{xx}(f)\right>_m = \frac1T\,\left<|X_T(f)|^2\right>_m
&&\text{(avg, $m$ spectra)}~.
\label{eqn:xsp-psd-avg}
\end{align}

As an obvious extension, the cross PSD of two generic random processes $\mathbf{x}(t)$ and $\mathbf{y}(t)$
\begin{align}
&S_{yx}(f) = 
\mathcal{F}\left\{R_{yx}(t')\right\}
\end{align}
is measured as
\begin{align}
&\left<S_{yx}(f)\right>_m = \frac1T\,\left<Y_T(f)\,X_T^\ast(f)\right>_m~.
\label{eqn:xsp-measured-Syx}
\end{align}

\subsection{Measurement time \boldmath$T$}
In practical experiments the measurement time is finite, so we can only access the truncated version $x_T(t)\leftrightarrow X_T(f)$ of a realization. 
In order to simplify the notation, the subscript $T$ for the \emph{truncation time} will be omitted.  Thus for example we write \req{eqn:xsp-measured-Syx} as
\begin{align*}
\left<S_{yx}(f)\right>_m&=\frac1T\,\left<Y(f)\,X^\ast(f)\right>_m
&&\text{(abridged notation)}~.
\end{align*}

\subsection{Why white Gaussian noise}
However too simplistic at first sight it may seem, the use of white Gaussian noise is justified as follows.  First, spectrally-smooth noise phenomena originate from large-number statistics (electrons and holes, semiconductor defects, shot noise, etc.), which by virtue of the central limit theorem yield to Gaussian process.  Second, most non-white noise phenomena of interest in follow the power-law model $S(f)=\sum h_if^i$, hence they can be converted into white noise after multiplication by a suitable power of $f$  without affecting the PDF, and converted back after analysis.  The idea of whitening and un-whitening a noise spectrum is by the way of far broader usefulness than shown here.  For these reasons we can take full benefit from the simplicity of white Gaussian noise.  Yet, it is understood that white noise rolls off at some point, so that all signals have finite power.

\section{The cross-spectrum method}
Recalling the definitions of Section \ref{sec:xsp-introduction}, we denote with $a(t)$ and $b(t)$ the background of the two instruments, with $c(t)$ the common noise, and with $A$, $B$ and $C$ their Fourier transform, letting the frequency implied.  Working with realizations, we no longer need a separate notation for the process.  By definition, $a(t)$, $b(t)$ and $c(t)$ are statistically independent.  We also assume that they are ergodic and stationary. The two instrument outputs are 
\begin{gather}
x(t)=c(t)+a(t)~~\leftrightarrow~~X = C+A\\
y(t)=c(t)+b(t)~~\leftrightarrow~~Y = C+B\makebox[0pt]{~~.}
\end{gather}

First, we observe that the cross-spectrum $S_{yx}$ converges to $S_{cc}$.  In fact, \begin{align}
\mathbb{E}\{S_{yx}\} 
&=\tfrac1T\,\mathbb{E}\{YX^\ast\}\nonumber\\
&=\tfrac1T\,\mathbb{E}\{[C+A]\times[C+B]^\ast\}\nonumber\\
&=\tfrac1T\,\bigl[\mathbb{E}\{CC^\ast\} + \mathbb{E}\{CB^\ast\} +
	\mathbb{E}\{AC^\ast\} + \mathbb{E}\{AB^\ast\}\bigr]\nonumber\\
&= S_{cc}
\end{align}
because the hypothesis of statistical independence gives
\begin{align*}
\mathbb{E}\{CB^\ast\}=0, \qquad
\mathbb{E}\{AC^\ast\}=0, \qquad\text{and}\qquad
\mathbb{E}\{AB^\ast\}=0~.
\end{align*}
Then we replace the expectation with the average on $m$ measured spectra
\begin{align}
\left<S_{yx}\right>_m 
&=\tfrac1T\,\left<YX^\ast\right>_m \nonumber\\
&=\tfrac1T\,\left<[C+A]\times[C+B]^\ast\right>_m\nonumber\\
&=\tfrac1T\,\bigl[\left<CC^\ast\right>_m + \left<CB^\ast\right>_m +
    \left<AC^\ast\right>_m + \left<AB^\ast\right>_m\bigr]\nonumber\\
&= S_{cc} + O(\sqrt{1/m})~, 
\label{eqn:xsp-syx-avg}
\end{align}
where $O(\:)$ means `order of.'  Owing to statistical independence, the cross terms decrease proportionally to $1/\sqrt{m}$.

\subsection{Statistical limit}
With no DUT noise it holds that $c=0$, hence $S_{cc}=0$.  
Maintaining the hypothesis of statistical independence of the two channels, we notice that the number of averaged spectra sets a statistical limit to the measurement.  
Only the cross terms remain in \req{eqn:xsp-syx-avg}, which decrease proportionally to $1/\sqrt{m}$.  Thus, the statistical limit is 
\begin{align}
\left<S_{yx}\right>_m &=  \tfrac1T\left<AB^\ast\right>_m
\approx\sqrt{\frac1m\,\left<S_{yy}\right>_m\left<S_{xx}\right>_m}
\qquad\text{(statistical limit)}.
\end{align}
Accordingly, a 5 dB improvement on the single-channel noise costs a factor of 10 in averaging, thus in measurement time.    The convergence law will be extensively discussed afterwards.

\subsection{Hardware limit}
Breaking the hypothesis of the statistical independence of the two channels, we are interested in the \emph{correlated noise} of the instrument, which limits the sensitivity.  This can be due for example to the crosstalk between the two channels, or to environmental fluctuations (ac magnetic fields, temperature, etc.) acting simultaneously on the two channels.  
The mathematical description is simplified by setting the true DUT noise to zero, and by re-interpreting $c(t)$ as the \emph{correlated noise} of the instrument observed on unlimited number of averaged spectra 
\begin{align}
\mathbb{E}\{S_{yx}\} = \mathbb{E}\{S_{cc}\}
\qquad\text{(hardware limit)}~.
\label{eqn:ddl-correl-hw-limit}
\end{align}
Nonetheless, the correct identification of this limit may require non-trivial experimental skill.

\subsection{Regular DUT measurement}
The accurate measurement of a regular DUT requires that 
\begin{enumerate}
\item The number $m$ is large enough for the statistical limit to be negligible
\item The hardware background noise is negligible as compared to the DUT noise
\end{enumerate}
In this conditions, the average cross spectrum converges to the expectation of the DUT noise
\begin{align}
\left<S_{yx}\right>_m ~~\rightarrow~~ \mathbb{E}\{S_{cc}\}
\qquad\text{(DUT measurement)}.
\label{eqn:xsp-correl-dut-meas-reg}
\end{align}
This is the regular use of the instrument.

\section{Running the experiment}\label{sec:fft-display}
Before getting through mathematical details, it is instructive to start from a simplified picture of what happens when we run an experiment.  For this purpose, we chose $\smash{\hat{S}}_{yx}=|\left<S_{yx}\right>_m|$ as an estimator of $S_{yx}$, which is often the default of the FFT analyzer in cross-spectrum mode.  This estimator is suitable to be displayed on a logarithmic scale (dB) because it takes only nonnegative values, but it is biased.
We observe the PSD on the display of the FFT analyzer as $m$ increases, looking for the signature of $\smash{\hat{S}}_{yx}$ converging to $S_{cc}$.

We restrict our attention to the case of DUT noise smaller than the single-channel background, as it usually occurs when we need the correlation.  The purpose for this  assumption is to make the simulations representative of the laboratory practice.  And of course we assume that the two channels are equal.

\subsection{Ergodicity}
\begin{figure}[t]
\centering\namedgraphics{scale=0.7}{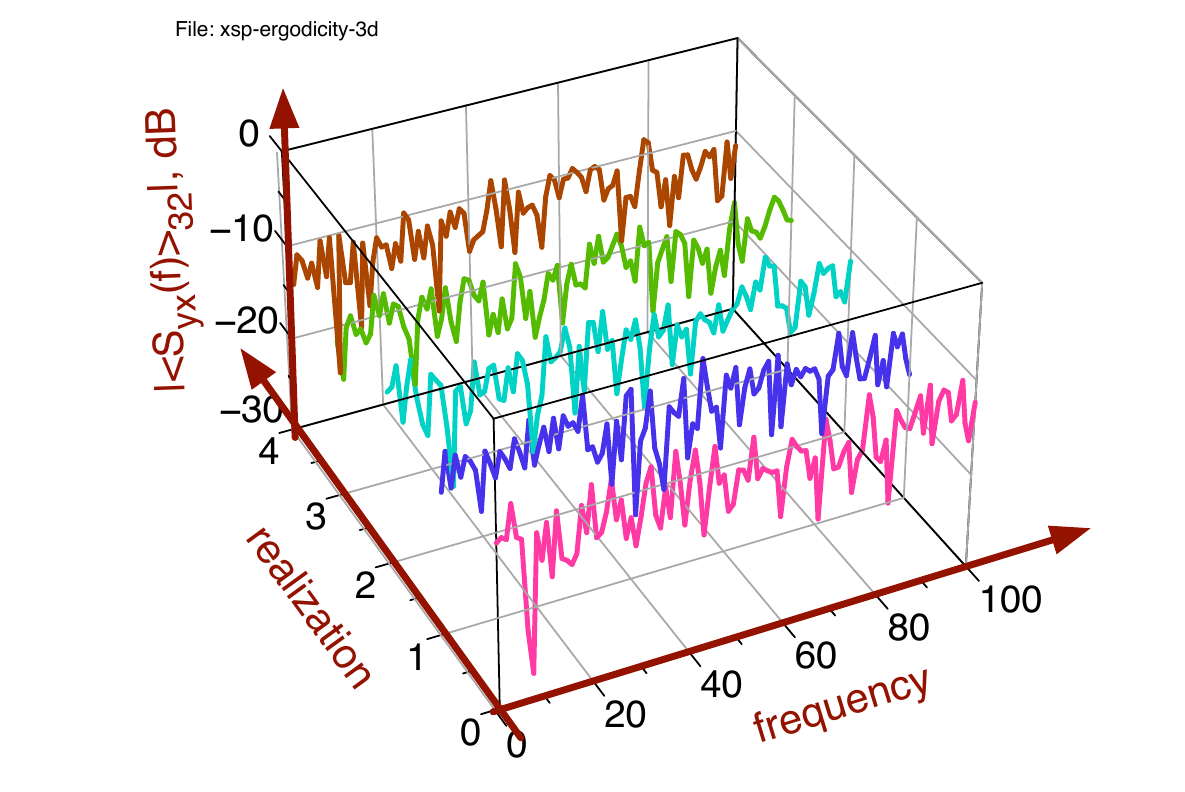}{\textwidth}
\caption{Sequence of cross spectra  $|\left<S_{yx}(f)\right>_{32}|$.}
\label{fig:xsp-ergodicity-3d}
\end{figure}
Averaging on $m$ realizations, the progression of a measurement gives a sequence of spectra $|\left<S_{yx}\right>_{m}|_i$ of running index $i$, as shown in Fig.~\ref{fig:xsp-ergodicity-3d}.
For a given frequency $f_0$, the sequence
$|\left<S_{yx}(f_0)\right>_{m}|_i$ is a time series.  
Since $S_{yx}(f_1)$ and $S_{yx}(f_2)$, are statistically independent for  $f_1\neq f_2$, also $|\left<S_{yx}(f_1)\right>_{m}|_i$ and $|\left<S_{yx}(f_2)\right>_{m}|_i$ are statistically independent.  For this reason, scanning the frequency axis gives access to (a subset of) the statistical ensemble.

\begin{figure}[t]
\centering\namedgraphics{scale=0.45}{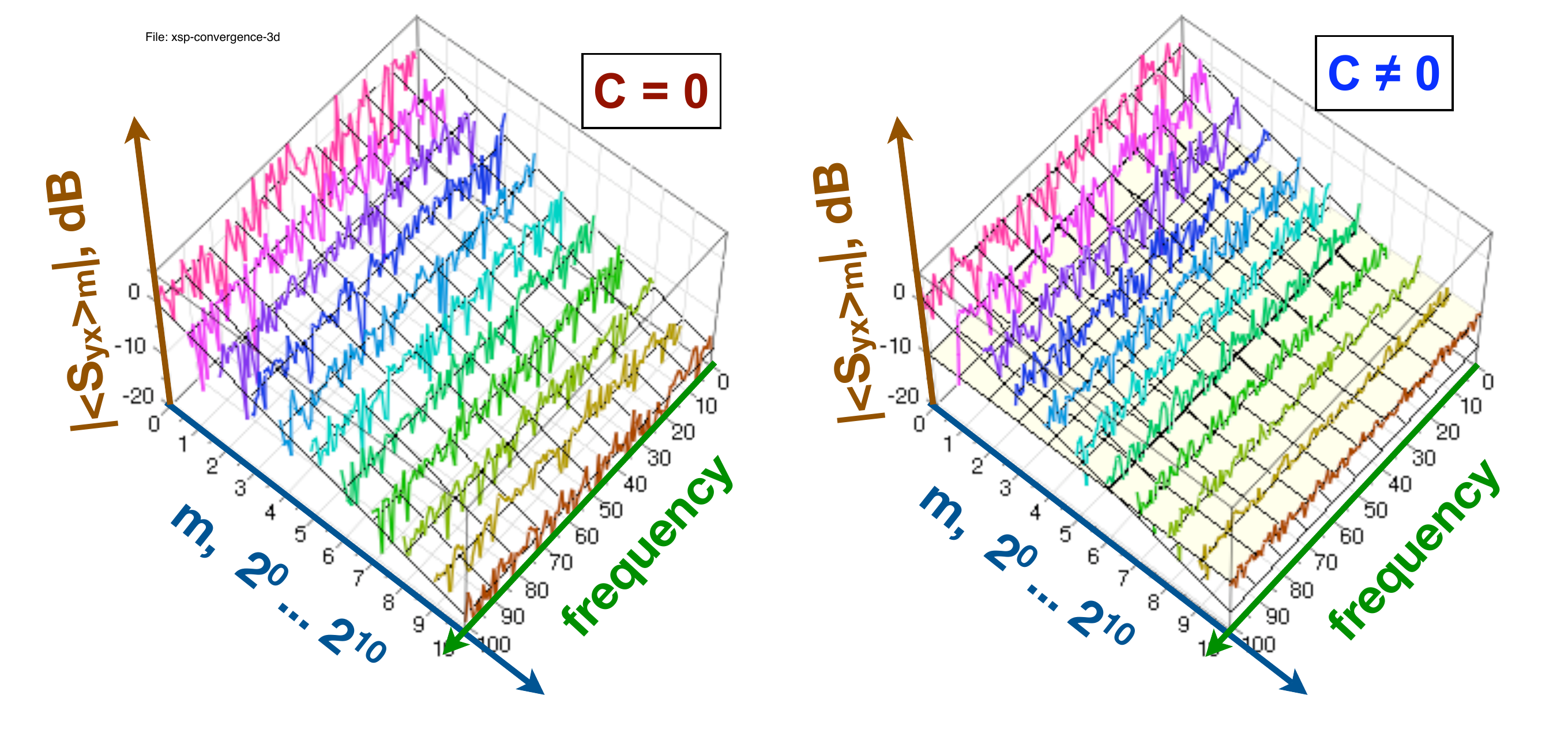}{\textwidth}
\caption{Sequence of cross spectra $|\left<S_{yx}\right>_{m}|$.}
\label{fig:xsp-convergence-3d}
\end{figure}
Ergodicity allows to interchange time statistics and ensemble statistics, thus the running index $i$ of the sequence and the frequency $f$.  The important consequence is that the average and the deviation calculated on the frequency axis give access to the average and deviation of the time series, without waiting for multiple realizations to be available.  This property helps detect when the cross spectrum leaves the $1/\sqrt{m}$ law and converges to the DUT noise.

Figure~\ref{fig:xsp-convergence-3d} shows a sequence of cross spectra $|\left<S_{yx}\right>_{m}|$, increasing $m$ in powers of two.  On the left-hand side of Fig.~\ref{fig:xsp-convergence-3d}, the DUT noise is set to zero.   Increasing $m$, the average cross spectrum decreases proportionally to $1/\sqrt{m}$, as emphasized by the slanted plane.  The $1/\sqrt{m}$ law is easily seen after averaging on the frequency axis separately for each value of $m$, and then transposing the law to each point of the frequency axis thanks to ergodicity. 
The right-hand side of Fig.~\ref{fig:xsp-convergence-3d} shows the same simulation, yet with the DUT noise set to a value of 10 dB lower than the single-channel background.  At small values of $m$ the cross-spectrum is substantially equal to the previous case.  Yet at $m\gtrsim100$ the cross-spectrum leaves the $1/\sqrt{m}$ law (slanted plane) and converges to the DUT noise (horizontal plane at $-10$ dB).  Once again, thanks to ergodicity we can transpose the average on the frequency axis to each point of the frequency axis.  

In the rest of this Section we will refer to a generic point of the PSD, letting the frequency unspecified.  The variable $f$ is omitted in order to simplify the notation.  Hence for example we will write  $\Re\{S_{yx}\}$ instead of $\Re\{S_{yx}(f)\}$.

\subsection{Single-channel noise.}
It is explained in Sec.~\ref{sec:xsp-estimation-Sxx} that the single-channel PSD $\left<S_{xx}\right>_m$ is $\chi^2$ distributed with $2m$ degrees of freedom.  The average PSD is equal to $\frac1T\,\mathbb{V}\{X\}=\frac1T\mathbb{V}\{A\} + \frac1T\mathbb{V}\{C\}$, where $\mathbb{V}\{\,\}$ is the variance;  the deviation-to-average ratio is equal to $1/\sqrt{m}$.  Of course the same holds for $S_{yy}$, after replacing $A$ with $B$.

\begin{figure}[t]
\centering\namedgraphics{scale=0.56, angle=90}{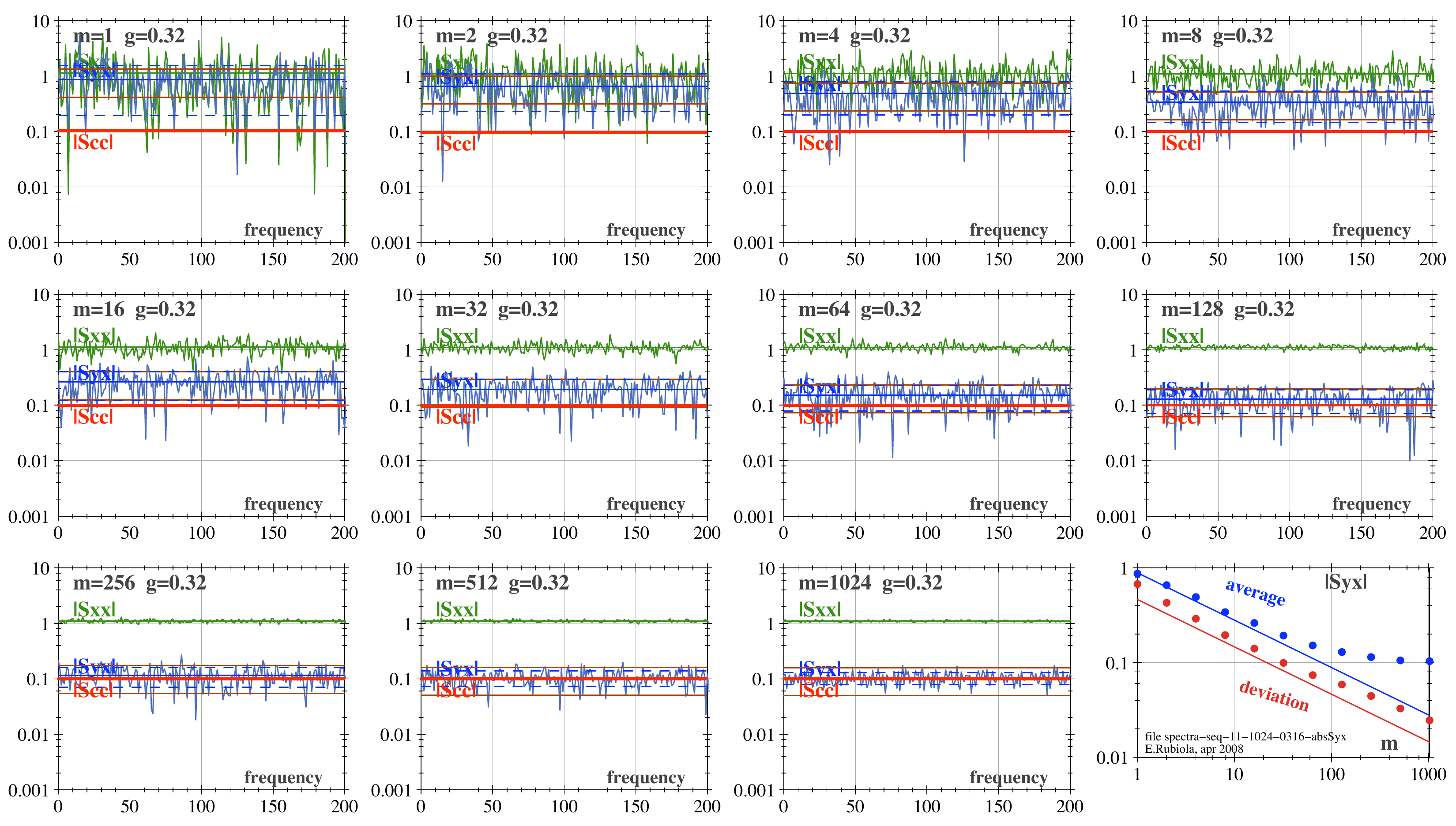}{\textwidth}
\caption{Simulated PSD, plotted for increasing number $m$ of averaged realizations.  The parameter $g=0.32$ ($-10$ dB), which is $\kappa$ in the main text, is the correlated noise, while the single-channel background is of one.}
\label{fig:spectra-seq-11-1024-0316-absSyx-WIDE}
\end{figure}

The track seen on the display converges to the DUT noise \emph{plus} the background noise, and shrinks as $m$ increases.  The track thickness is twice the deviation.  This fact is shown on Fig.~\ref{fig:spectra-seq-11-1024-0316-absSyx-WIDE}.   The green plot, labeled $|S_{xx}|$, keeps the same vertical position as $m$ increases, and shrinks.

\subsection{Cross-spectrum observed with insufficient \boldmath$m$.}  
When the number $m$ of averaged realizations is insufficient for the DUT noise to show up, the system behaves as the two channels were (almost) statistically independent.  In this conditions we can predict the spectrum by setting $X\simeq A$, $Y\simeq B$ and $C\simeq0$, thus $\mathbb{E}\{S_{yx}\}\simeq0$.  

The estimator $\smash{\hat{S}}_{yx}=|\left<S_{yx}\right>_m|$ has Rayleigh distribution with $2m$ degrees of freedom.  
Normalizing on the single-channel background $\mathbb{E}\{S_{xx}\}=\mathbb{E}\{S_{yy}\}=1$, and using the results of Sec.~\ref{sec:xsp-noise-rejection}, we find that
\begin{align}
\mathbb{E}\{\hat{S}_{yx}\}
&=\mathbb{E}\{|\left<S_{yx}\right>_m|\}=\sqrt{\frac{\pi}{4m}}=\frac{0.886}{\sqrt{m}}\nonumber\\[1ex]
\mathbb{V}\{\hat{S}_{yx}\}
&=\mathbb{V}\{|\left<S_{yx}\right>_m|\}
=\frac1m\left(1-\frac{\pi}{4}\right)=\frac{0.215}{m}~,\nonumber
\intertext{and therefore}
\mathrm{dev}\{\hat{S}_{yx}\}
&=\sqrt{\mathbb{V}\{|\left<S_{yx}\right>_m|\}}
=\sqrt{\frac{1}{m}\left(1-\frac{\pi}{4}\right)}=\frac{0.463}{\sqrt{m}}\nonumber\\[1ex]
\frac{\mathrm{dev}\{\hat{S}_{yx}\}}{\mathbb{E}\{\hat{S}_{yx}\}}
&=\sqrt{\frac{4}{\pi}-1}=0.523\qquad\text{(independent of $m$)}~.\nonumber
\end{align}
The track is centered at $\smash{\frac{0.886}{\sqrt{m}}}$.  This is the estimator bias.  The track looks as a horizontal band located at $\mathrm{avg}\pm\mathrm{dev}$, thus on a logarithmic from $10\log_{10}(1-\mathrm{dev/avg})=-3.21~\unit{dB}$ to $10\log_{10}(1+\mathrm{dev/avg})=+1.83~\unit{dB}$ asymmetrically distributed around the average.  This is shown on Fig.~\ref{fig:spectra-seq-11-1024-0316-absSyx-WIDE}.   For $m\lesssim100$, the blue plot labeled $|S_{yx}|$ decreases proportionally to $1/\sqrt{m}$ and has the constant thickness of half a decade (5 dB), independent of $m$.

\subsection{Cross-spectrum observed with large \boldmath$m$.}  When the number $m$ of averaged realizations is large enough, the background noise vanishes and the DUT spectrum shows up.  The cross spectrum no longer decreases but the variance still does.  Qualitatively speaking, the average is set by the DUT noise $S_{cc}$ and the deviation is set by the instrument background divided by $\sqrt{m}$.  On a logarithmic scale, the track no longer decreases and starts shrinking.  This is shown on Fig.~\ref{fig:spectra-seq-11-1024-0316-absSyx-WIDE} for $m\gtrsim100$, blue plot labeled $|S_{yx}|$.

The above reasoning can be reversed.  The simultaneous observation that the cross spectrum \emph{stops decreasing}, and \emph{shrinks} is the signature that the averaging process is converging.  The single-channel background is rejected and the instrument measures the DUT noise (or the hardware limit, which is higher).  This fact is of paramount importance in some measurements, where for some reasons we cannot remove the DUT.

\section{Estimation of \boldmath$S_{xx}$}\label{sec:xsp-estimation-Sxx}
The measurement accuracy depends on three main factors, instrument calibration, instrument background (front-end and quantization), and statistical estimation.  Only the latter is analyzed in this Section.

\begin{figure}[t]
\centering\namedgraphics{scale=0.6}{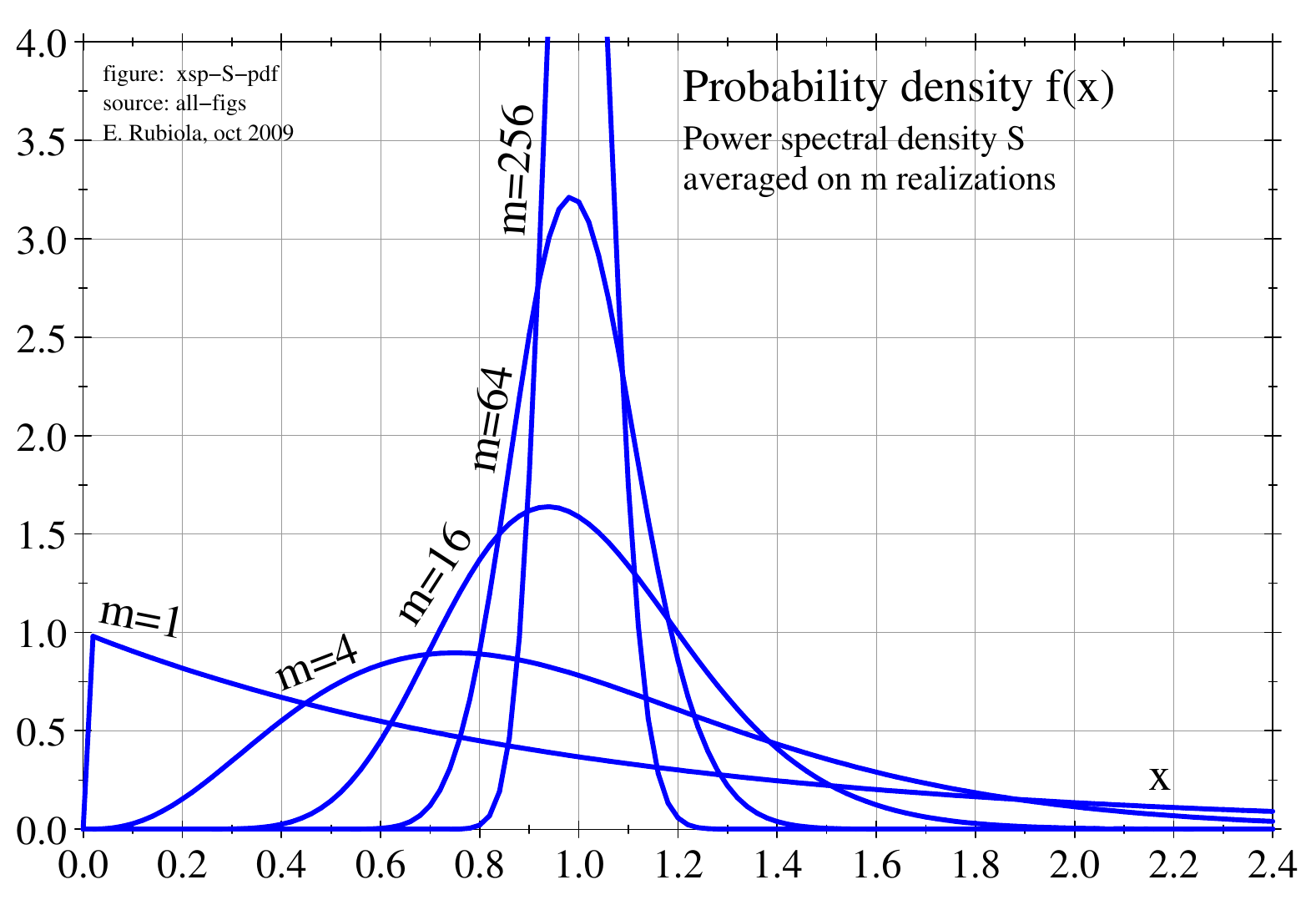}{\textwidth}
\caption{Probability density function $f(x)$ of the PSD averaged on $m$ realizations.}
\label{fig:xsp-S-pdf}
\end{figure}

As a property of zero-mean white Gaussian noise, the Fourier transform $X=X'+\imath X''$ is also zero-mean Gaussian, and the energy is equally split between $X'$ and $X''$.
Restricting our attention to a generic point (i.e., to an unspecified frequency), the PSD is
\begin{align*}
\mathbb{E}\{S_{xx}\} &= \frac1T\,\mathbb{E}\Bigl\{\left|X\right|^2\Bigr\} 
	= \frac1T\,\mathbb{E}\Bigl\{\left[X'^{\,2}+X''^{\,2}\right]\Bigr\}~.
\end{align*}
For use in this Section we define  
\begin{align*}
\varsigma^2=\mathbb{E}\{S_{xx}\}~, 
\end{align*} 
which is the power in 1 Hz bandwidth.
Since $X'$ and $X''$ are zero-mean Gaussian-distributed random variables, a single realization 
\begin{align*}
S_{xx} &= \frac1T\,\left[X'^{\,2}+X''^{\,2}\right]
\end{align*}
follows a  $\chi^2$ distribution with two degrees of freedom.  
After our definition of $\varsigma^2$, we find that
\begin{align*}
\mathbb{V}\{X'\}=\mathbb{V}\{X''\}=\frac{T}{2}\,\varsigma^2~.
\end{align*}
because $S_{xx}$ includes a factor $\frac1T$.
This is seen on the ``scaled $\chi^2$'' column of Table~\ref{tab:xsp-chi-square-prop}, after setting $\nu=2$ (degrees of freedom) and $\sigma=\frac{1}{2}T\,\varsigma^2$.  On that Table we find that $\mathbb{E}\{S_{xx}\}=\frac{1}{T}\,\nu\sigma^2$, which is equal to $\varsigma^2$, and that $\mathbb{V}\{S_{xx}\}=\frac{1}{T^2}\,2\nu\sigma^4$, hence 
\begin{align*}
\mathbb{V}\{S_{xx}\}=\varsigma^4~.
\end{align*}
Averaging on $m$ realizations of $S_{xx}$
\begin{align*}
\left<S_{xx}\right>_m = \frac{1}{m} \sum_{i=1}^{m} \;\frac1T\left[X_i'^{\,2}+X_i''^{\,2}\right], 
\end{align*}
we notice that $\left<S_{xx}\right>_m$ has $\chi^2$ distribution with $2m$ degrees of freedom.  Using the right-hand column of Table~\ref{tab:xsp-chi-square-prop}, we find $\mathbb{V}\{\left<S_{xx}\right>_m\}=\frac1m\varsigma^4$.
The uncertainty (standard deviation) is therefore
\begin{align*}
\text{dev}\{\left<S_{xx}\right>_m\} &= \frac{1}{\sqrt{m}} \varsigma^2
&\frac{\text{dev}\{\left<S_{xx}\right>_m\} }{\mathbb{E}\{\left<S_{xx}\right>_m\}} &= \frac{1}{\sqrt{m}}~.
\end{align*}
Figure \ref{fig:xsp-S-pdf} shows an example PDF of the spectrum averaged on $m$ realizations.  The $\chi^2$ distribution is normalized for the standard deviation to be equal one.  Increasing $m$, the PDF converges to the normal distribution and shrinks.

Finally, we may find useful the following normalization
\begin{align*}
S_{aa}&=1\quad\text{(background)} & S_{cc}&=\kappa^2\quad\text{(DUT)}~.
\end{align*}
Expanding $X=X'+\imath X''=(A'+C')+\imath(A''+C'')$ we notice that $X$ is zero-mean white Gaussian noise, and that
\begin{align*}
\mathbb{E}\left\{\left<S_{xx}\right>_m\right\}&=1+\kappa^2 &
\text{dev}\left\{\left<S_{xx}\right>_m\right\}=\frac{1+\kappa^2}{\sqrt{m}}~.
\end{align*}

\section{Estimation of \boldmath$S_{yx}$ and noise rejection}\label{sec:xsp-noise-rejection}
It is obvious from Eq.~\req{eqn:xsp-psd-avg} that the spectrum $S_{xx}(f)$ takes always \emph{real positive} values, even if averaged on a small number of realizations.  Since some kind of fundamental noise is always present in a physical experiment, the probability that $S_{xx}(f)$ nulls at some frequency is zero.  Conversely, the cross-spectrum $S_{yx}(f)$ is a \emph{complex} function that converges to the positive function $S_{cc}(f)$ only after averaging on a sufficient number $m$ of realizations, as seen in Eq. \req{eqn:xsp-syx-avg}.

In numerous practical cases we need to plot $S_{yx}(f)$ on a logarithmic vertical scale, which is of course impossible where $S_{yx}(f)$ is not positive.  
\begin{itemize}
\item In radio engineering virtually all spectra are given in decibels, which resorts to a logarithmic scale.
\item When the spectrum spreads over a large dynamic range, only a compressed scale makes sense.  The logarithmic scale is by far the preferred representation.
\item Numerous spectra found in physical experiments follow a polynomial law because the time-domain derivative (integral) maps into a multiplication (division) of the spectrum by $f^2$.  On a logarithmic plot, a power of $f$ maps into a straight line.
\item  It is explained in Section \ref{sec:fft-display} that running the experiment, average and deviation of the instrument noise are ruled by the same $1/sqrt{m}$ law until the number of averaged realizations is sufficient for $S_{yx}(f)$ to converge to $S_{cc}(f)$.  This is most comfortably seen on a logarithmic scale.
\end{itemize}
Thus, we need to extend Section \ref{sec:xsp-estimation-Sxx} to the cross spectrum, discussing the suitable estimators.  The estimator may introduce noise and bias.  In everyday life a better estimator may save only a little amount of time, and in this case it could be appreciated mainly because it is smarter.  Oppositely in long-term measurements, like \emph{timekeeping} and \emph{radioastronomy}, a single data point takes years of observation.  Here, the choice of the estimator may determine whether the experiment is feasible or not.

\subsection{Basic material}
Let us expand $S_{yx}$
\begin{align}
S_{yx}
&=\tfrac1T\,\mathbb{E}\left\{YX^\ast\right\}\nonumber\\
&=\tfrac1T\,\mathbb{E}\left\{(B+C)\times(A+C)^\ast\right\}\nonumber\\
&=\tfrac1T\,\mathbb{E}\left\{(B'+\imath B''+C'+\imath C'')\times(A'-\imath A''+C'-\imath C'')\right\}\nonumber\\
&=\tfrac1T\,\mathbb{E}\left\{\bigl(B'A'+B''A''+B'C'+B''C''+C'A'+C''A''+C'^{\,2}+C''^{\,2}\bigr)\right.\nonumber\\[0.5ex]
&\left.\qquad+\imath\bigl(B''A'-B'A''+B''C'-B'C''+C''A'-C'A''\bigr)\right\}
\label{eqn:xsp-correl-dut-meas}
\end{align}
and simplify the calculus by normalizing on the variances as follows
\begin{align}
\mathbb{V}\{A\}&=1 & \mathbb{V}\{A'\}&=1/2 &  \mathbb{V}\{A''\}&=1/2\nonumber\\
\mathbb{V}\{B\}&=1 &  \mathbb{V}\{B'\}&=1/2 &  \mathbb{V}\{B''\}&=1/2\nonumber\\
\mathbb{V}\{C\}&=\kappa^2\ll1  &  \mathbb{V}\{C'\}&=\kappa^2/2 &  \mathbb{V}\{C''\}&=\kappa^2/2~.\nonumber
\end{align}
Notice that an additional factor $T$ must be added a-posteriori for a proper normalization on $\mathbb{E}\{S_{aa}\}=\mathbb{E}\{S_{bb}\}=1$ (background power  in 1 Hz bandwidth equal to one), as we did in Section \ref{sec:xsp-estimation-Sxx}.  Thanks to energy equipartition, it follows that $\mathbb{V}\{A'\}=1/2\Rightarrow\mathbb{V}\{A'\}=T/2$, etc.

The assumption that $\kappa^2\ll1$, though not necessary, is quite representative of actual experiments because the main virtue of the correlation method is the capability of extracting the DUT noise when it is lower than the background.
 
Looking at \req{eqn:xsp-correl-dut-meas}, we identify the following classes
\def\pba#1{\parbox{27ex}{\setlength{\baselineskip}{2.5ex}#1}}
\def\pbb#1{\parbox{22ex}{\setlength{\baselineskip}{2.5ex}#1}}
\def\rulethickness{0pt}
\begin{center}
\begin{tabular}{|l|c|c|c|l|}\hline
\rule[-1.5ex]{\rulethickness}{4ex}%
terms & $\mathbb{E}$ & $\mathbb{V}$ & PDF & comment\\\hline
\rule[-2.5ex]{\rulethickness}{6ex}%
\pba{$B'A'$, $B''A''$, $B''A'$, $B'A''$}&0&$1/4$&Gauss
&\pbb{product of zero-mean Gaussian processes}\\\hline
\rule[-2.5ex]{\rulethickness}{6ex}%
\pba{$B'C'$, $B''C''$, $C'A'$, $C''A''$	, $B''C'$, $B'C''$, $C''A'$, $C'A''$}
&0&$\kappa^2/4$&Gauss
&\pbb{product of zero-mean Gaussian processes}\\\hline
\rule[0ex]{\rulethickness}{2.5ex}%
$C'^{\,2}+C''^{\,2}$&$\kappa^2$&$\kappa^4$&$\chi^2$&sum of zero-mean\\
				&&&$\nu=2$& square Gaussian proc.\\\hline
\end{tabular}
\end{center}
Equation \req{eqn:xsp-correl-dut-meas} can be rewritten as 
\begin{align}
S_{yx}&=\tfrac1T\,\mathbb{E}\left\{\mathscr{A}+\imath\mathscr{B}+\mathscr{C}\right\}
\label{eqn:xsp-correl-dut-meas-ABC}
\intertext{where the terms}
\mathscr{A}&=B'A'+B''A''+B'C'+B''C''+C'A'+C''A''\nonumber\\
\mathscr{B}&=B''A'-B'A''+B''C'-B'C''+C''A'-C'A''\nonumber\\
\mathscr{C}&=C'^{\,2}+C''^{\,2}\nonumber
\end{align}
have the statistical properties listed underneath.  Notice that $\left<\mathscr{C}\right>_m$ follows a $\chi^2$ distribution with $2m$ degrees of freedom, thus for large $m$ it can be approximated with a Gaussian distributed variable of equal average and variance, which is denoted with $\bigl<\tilde{\mathscr{C}}\bigr>_m$.
\def\rulethickness{0pt}
\begin{center}
\begin{tabular}{|l|c|c|c|l|}\hline
\rule[-1.5ex]{\rulethickness}{4ex}%
term & $\mathbb{E}$ & $\mathbb{V}$ & PDF & comment\\\hline
\rule[-2.5ex]{\rulethickness}{6.5ex}%
$\left<\mathscr{A}\right>_m$&0&$\displaystyle\frac{1+2\kappa^2}{2m}$&Gauss&average (sum) of zero-mean\\\cline{1-4}
\rule[-2.5ex]{\rulethickness}{6.5ex}%
$\left<\mathscr{B}\right>_m$&0&$\displaystyle\frac{1+2\kappa^2}{2m}$&Gauss&Gaussian processes\\\hline
\rule[0ex]{\rulethickness}{2.5ex}%
$\left<\mathscr{C}\right>_m$&$\kappa^2$&$\displaystyle\kappa^4/m$&$\chi^2$&average (sum) of\\
				&&&$\nu=2m$& chi-square processes\\\hline
\rule[-1.5ex]{\rulethickness}{4ex}%
$\bigl<\tilde{\mathscr{C}}\bigr>_m$&$\kappa^2$&$\displaystyle\kappa^4/m$&Gauss& approximates $\left<\mathscr{C}\right>_m$ for large $m$\\\hline
\end{tabular}
\end{center}
Next, we will analyze the properties of some useful estimators of $\hat{S}_{yx}$.
Running an experiment, the logarithmic plot is comfortable because the average-to-deviation ratio is easily identified as the thickness of the track, independent of the vertical position.  Yet, the logarithmic plot can only be used to display nonnegative quantities.

\subsection{\boldmath$\hat{S}_{yx}=\left|\left<S_{yx}\right>_m\right|$}\label{ssec:xsp-estimator-Abs-value}
The main reason for us to spend attention with this estimator is that it is the default setting for cross-spectrum measurement in most FFT analyzers.  Besides, it can be used in conjunction with $\arg\left<S_{yx}\right>_m$ when the hypothesis that the delay of the two channels is not equal and useful information is contained in the argument, as it happens in radio-astronomy.  $|\left<S_{yx}\right>_m|$ is of course suitable to logarithmic plot because it can only take nonnegative values.
The relevant objections against this estimator are
\begin{itemize}
\item There is no need to take in $\Im\left\{S_{yx}\right\}$, which contains half of the total background noise.
\item The instrument background turns into relatively large estimation bias.
\end{itemize}
For large $m$, where $\left<\mathscr{C}\right>_m$ tends to $\left<\smash{\tilde{\mathscr{C}}}\right>_m$, the estimator is expanded as
\begin{align*}
|\left<S_{yx}\right>_m| 
&=\frac1T\sqrt{\left[\Re\left\{\left<YX^\ast\right>_m\right\} \right]^2 
	+ \left[\Im\left\{\left<YX^\ast\right>_m\right\} \right]^2}\\
&=\frac1T \sqrt{\left[\left<\mathscr{A}\right>_m+\left<\smash{\tilde{\mathscr{C}}}\right>_m \right]^2 
	+ \left[\left<\mathscr{B}\right>_m\right]^2}~.
\end{align*}

\subsubsection{The (not so) silly case of \boldmath$\kappa=0$}
\begin{figure}[t]
\centering\namedgraphics{scale=0.6}{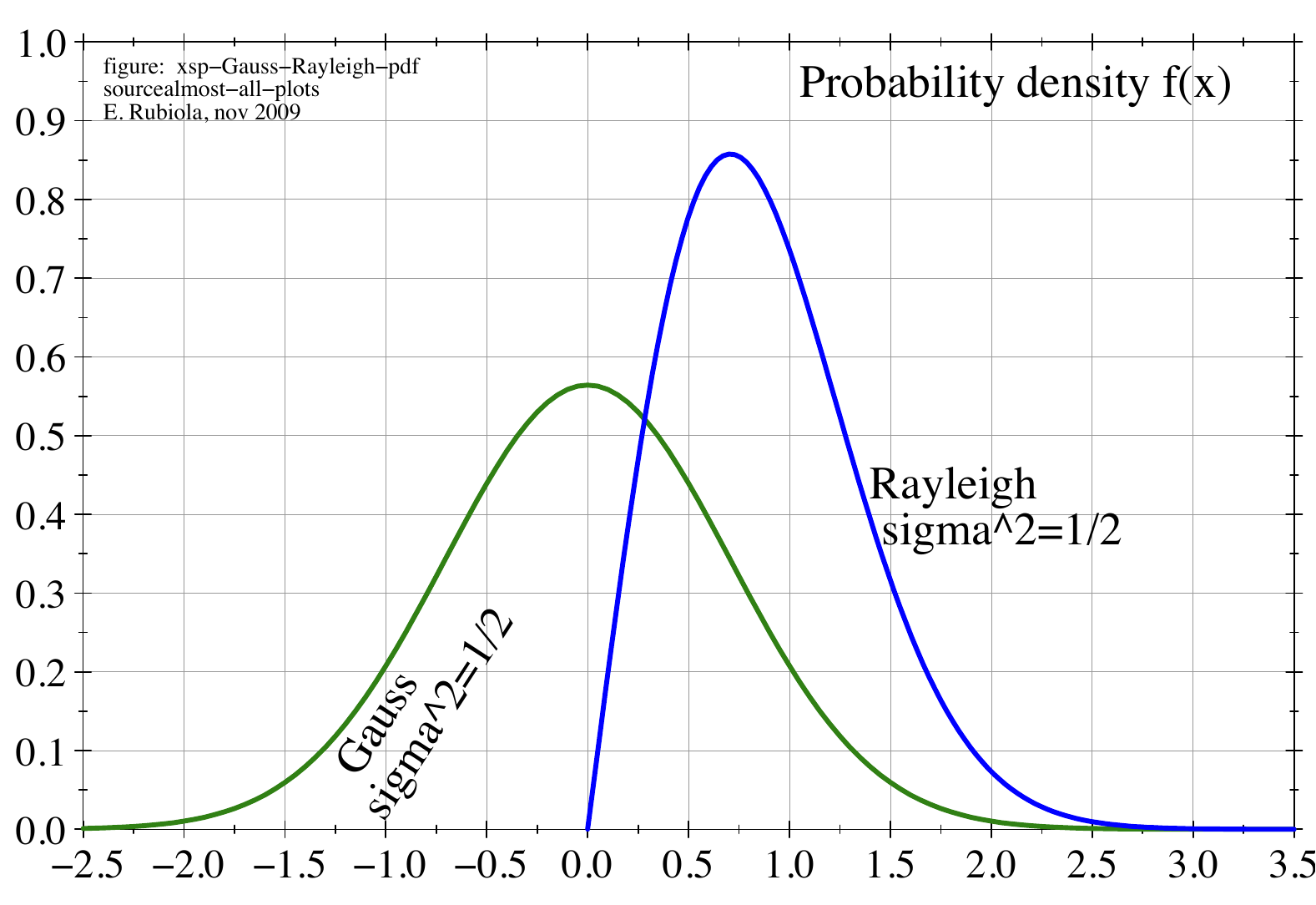}{\textwidth}
\caption{Gaussian distribution of variance $\sigma^2=1/2$ and Rayleigh distribution generated by a pair of Gaussian variables of variance $\sigma^2=1/2$.}
\label{fig:xsp-Gauss-Rayleigh-pdf}
\end{figure}
The analysis of this case tells us what happens when $m$ is insufficient for the single-channel to be rejected, so that the displayed average spectrum is substantially the bias of the estimator.
Since $c\leftrightarrow C=0$, it holds that $\mathscr{C}=0$.  Letting 
\begin{align*}
\left<\mathscr{Z}\right>_m&=\sqrt{\left[\left<\mathscr{A}\right>_m\right]^2 
	+ \left[\left<\mathscr{B}\right>_m\right]^2}~.
\end{align*}
we notice that $\left<\mathscr{Z}\right>_m$ is Rayleigh distributed with $2m$ degrees of freedom.  
Using Table~\ref{tab:xsp-rayleigh}, we find that
\begin{align*}
&\mathbb{E}\{\left<\mathscr{Z}\right>_m\}=\sqrt{\frac{\pi}{4m}}=\frac{0.886}{\sqrt{m}}&&\text{(average)}\\[1ex]
&\mathbb{V}\{\left<\mathscr{Z}\right>_m\}
=\frac1m\left(1-\frac{\pi}{4}\right)=\frac{0.215}{m}&&\text{(variance)}
\end{align*}
Figure~\ref{fig:xsp-Gauss-Rayleigh-pdf} compares the case $m=1$ (Rayleigh distribution) to the Gaussian distribution associated with the best estimator (Section~\ref{ssec:xsp-estimator-Real-part}).

Interestingly, the deviation-to-average ratio, which also applies to $|\left<S_{yx}\right>_m|$,
\begin{align}
\frac{\displaystyle\text{dev}\{|\left<S_{yx}\right>_m|\}}{\displaystyle\mathbb{E}\{|\left<S_{yx}\right>_m|\}}
&=\sqrt{\frac{4}{\pi}-1}=0.523
\qquad\frac{\text{dev}}{\mathbb{E}}
\end{align}
is independent of $m$.
In logarithmic scale, the cross spectrum appears as a strip decreasing as $5\log(m)$ dB, yet \emph{of constant thickness} of approximately 5 dB (dev/avg).  This is seen in the example of Fig.~\ref{fig:spectra-seq-11-1024-0316-absSyx-WIDE}.

\subsubsection{Large number of averaged realizations}
The estimator converges to $\kappa^2$, which is trivial, and for $\kappa\ll1$ the deviation-to-average ratio is approximately $1/\sqrt{m}$.  This issue is not further expanded here.

\subsection{\boldmath$\hat{S}_{yx}=\Re\left\{\left<S_{yx}\right>_m\right\}$}\label{ssec:xsp-estimator-Real-part}
\begin{figure}[t]
\centering\namedgraphics{scale=0.8}{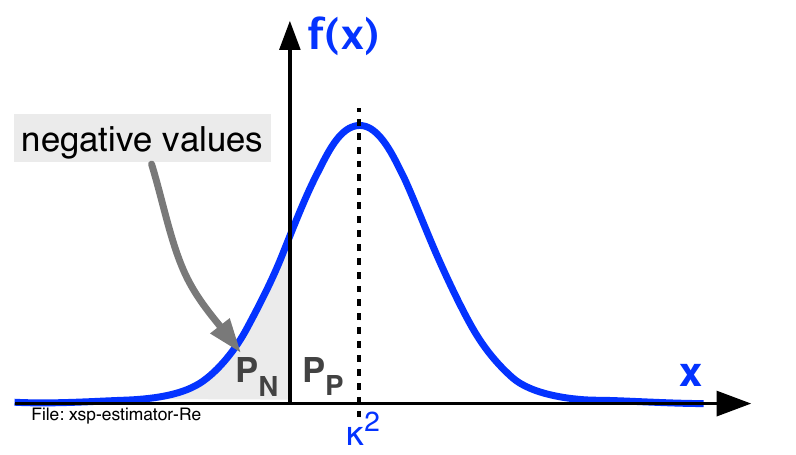}{\textwidth}
\caption{PDF of the estimator $\hat{S}_{yx}=\Re\left\{\left<S_{yx}\right>_m\right\}$.}
\label{fig:xsp-estimator-Re}
\end{figure}

This is the best estimator to the extent that
\begin{itemize}
\item All the useful information is in $\Re\left\{S_{yx}\right\}=\frac1T(\mathscr{A}+\mathscr{C})$.
\item Since the instrument background is equally split in $\Re\left\{S_{yx}\right\}$ and $\Im\left\{S_{yx}\right\}$, discarding $\Im\left\{S_{yx}\right\}$ results in 3 dB improvement of the SNR\@.
\item for the same reason, the instrument background does not contribute to the bias.
\end{itemize}
The main drawback is that this estimator is not suitable to logarithmic plot because $\Re\left\{\left<S_{yx}\right>_m\right\}$ can take negative values, especially at small $m$.  
For large $m$ we can approximate $\left<\mathscr{C}\right>_m$ with $\bigl<\smash{\tilde{\mathscr{C}}}\bigr>_m$, which is Gaussian distributed.  
Letting 
\begin{align*}
\left<\mathscr{Z}\right>_m=\left<\mathscr{A}\right>_m+\left<\smash{\tilde{\mathscr{C}}}\right>_m~, 
\end{align*}
the PDF of $\left<\mathscr{Z}\right>_m$ is Gaussian (Fig.~\ref{fig:xsp-estimator-Re}).
Using the results of Sec.~\ref{ssec:xsp-gaussian}, we find
\begin{align}
\mathbb{E}\left\{\left<\mathscr{Z}\right>_m\right\}&=\kappa^2\\
\mathbb{V}\left\{\left<\mathscr{Z}\right>_m\right\}&=\frac{1+2\kappa^2+2\kappa^4}{2m}\\
\text{dev}\left\{\left<\mathscr{Z}\right>_m\right\}&=\sqrt{\frac{1+2\kappa^2+2\kappa^4}{2m}}
\approx\frac{1+\kappa^2}{\sqrt{2m}}\\
\frac{\text{dev}\left\{\left<\mathscr{Z}\right>_m\right\}}{%
\mathbb{E}\left\{\left<\mathscr{Z}\right>_m\right\}}
&=\frac{\sqrt{1+2\kappa^2+2\kappa^4}}{\kappa^2\:\sqrt{2m}}
\approx\frac{1+\kappa^2}{\kappa^2\:\sqrt{2m}}
\label{eqn:xsp-dev-avg-Re}\\[1ex]
P_N&=\frac12 \text{erfc}\!\left(\frac{\kappa^2}{\sqrt{2}\:\sigma}\right)
	&&(\mathbb{P}\{\mathbf{x}<0\},~\text{Sec.~\ref{ssec:xsp-gaussian}})\\
P_P&=1-\frac12 \text{erfc}\!\left(\frac{\kappa^2}{\sqrt{2}\:\sigma}\right)
	&&(\mathbb{P}\{\mathbf{x}>0\},~\text{Sec.~\ref{ssec:xsp-gaussian}})~.
\end{align}
Accordingly, for $\kappa\ll1$ a 0 dB SNR requires that $m=\frac{1}{2\kappa^4}$.  If for example the DUT noise is 20 dB lower than the single-channel background, thus $\kappa=0.1$, averaging on $5{\times}10^3$ spectra is necessary to get a SNR of 0 dB.
On the other hand, if $\kappa\gg1$ the deviation-to-average ratio converges to $1/\sqrt{2m}$, which is what we expect if the instrument background is negligible.

\subsubsection{Precision vs.\ energy conservation}
The term $\sqrt{2}$ in the denominator of \req{eqn:xsp-dev-avg-Re} means that the SNR of the correlation system is 3 dB better than the single-channel system.  In a physical system ruled by energy conservation this factor does not come for free because the DUT power is equally split into two channels.  The conclusion is that the factor $\sqrt{2}$ in the SNR cancels with the $\sqrt{2}$ intrinsic loss of the power splitter.  So, the basic \emph{conservation laws} of thermodynamics (or information) are \emph{not violated}.

\subsection{\boldmath$\hat{S}_{yx}=\left|\Re\left\{\left<S_{yx}\right>_m\right\}\right|$}\label{ssec:xsp-estimator-Abs-Real-part}
\begin{figure}[t]
\centering\namedgraphics{scale=0.8}{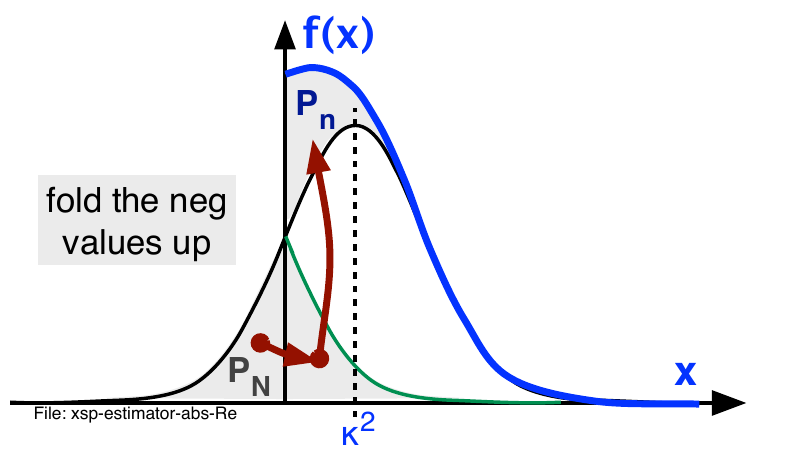}{\textwidth}
\caption{PDF of the estimator $\hat{S}_{yx}=\left|\Re\left\{\left<S_{yx}\right>_m\right\}\right|$.}
\label{fig:xsp-estimator-abs-Re}
\end{figure}
The negative values of $\left<S_{yx}\right>_m$ are folded up, so that $\smash{\hat{S}}_{yx}$ is always positive and can be plotted on a logarithmic axis.
Approximating $\left<\mathscr{C}\right>_m$ with $\left<\smash{\tilde{\mathscr{C}}}\right>_m$ for large $m$, the estimator is expanded as
\begin{align*}
\left|\Re\left\{\left<S_{yx}\right>_m\right\}\right|
&=\frac1T \left| \left<\mathscr{A}\right>_m +\left<\smash{\tilde{\mathscr{C}}}\right>_m\right|
\end{align*}
The PDF of $|\Re\{\left<S_{yx}\right>_m\}|$ is obtained from the PDF of $|\Re\{\left<S_{yx}\right>_m\}|$ already studied in Section~\ref{ssec:xsp-estimator-Real-part} by folding%
\footnote{A theorem states that follows.  Let $\mathbf{x}$ a random variable, $f(x)$ its PDF, and $\mathbf{y}=|\mathbf{x}|$ a function of $\mathbf{x}$.  The PDF of $\mathbf{y}$ is $g(y)=f(y)\mathfrak{u}(y)+f(-y)\mathfrak{u}(-y)$, where $\mathfrak{u}(y)$ is the Heaviside (step) function.  Notice that the term $f(-y)\mathfrak{u}(-y)$ is the negative-half-plane ($y<0$) side of $f(y)$ folded to the positive half plane.}
the negative-half-plane of the original PDF on the positive half plane.
The result is shown in Fig.~\ref{fig:xsp-estimator-abs-Re}.

\subsection{\boldmath$\hat{S}_{yx}=\Re\left\{\left<S_{yx}\right>_{m'}\right\}$, averaging on the positive values}%
\label{ssec:xsp-estimator-Neg-discarded} 
\begin{figure}[t]
\centering\namedgraphics{scale=0.8}{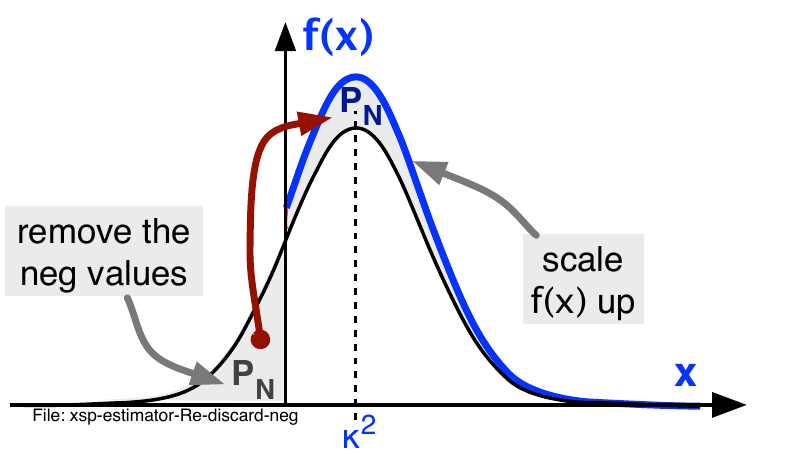}{\textwidth}
\caption{PDF of the estimator obtained averaging the positive values of $\Re\left\{S_{yx}\right\}$.}
\label{fig:xsp-estimator-Re-discard-neg}
\end{figure}
Averaging $m$ values of $\Re\{S_{yx}\}$, we expect $m'=m\,P_P$ positive values and $m-m'=m\,P_N$ negative values.  This estimators consists of averaging on the $m'$ positive values, discarding the negative values.
As usual, we assume that for large $m$ the term $\left<\mathscr{C}\right>_m$ is approximated with $\left<\smash{\tilde{\mathscr{C}}}\right>_m$, so that its PDF is Gaussian.
The PDF of this estimator is formed%
\footnote{A theorem states that follows.  Let $f(x)$ the PDF of a process, and $g(x)$ the PDF conditional to the event $\mathbf{e}$.  The conditional PDF is obtained in two steps.  First an auxiliary function $h(x)$ is obtained from $f(x)$ by selecting the sub-domain defined by $\mathbf{e}$.  Second, the desired PDF is $g(x)=h(x)/\int_{-\infty}^{\infty}h(x)\:dx$.  The first step generates $h(x)$ equal to $f(x)$, but taking away the portions not allowed by $\mathbf{e}$.  The second step scales the function $h(x)$ up so that $\int_{-\infty}^{\infty}g(x)\:dx=1$ (probability of all possible events), thus it is a valid PSD.} 
from the PDF of $\Re\{\left<S_{yx}\right>_m\}$ after removing the negative-half-plane values and scaling up the result for the integral of the PDF to be equal to one.  This is illustrated in Fig.~\ref{fig:xsp-estimator-Re-discard-neg}.

\subsection{Estimator \boldmath$\hat{S}_{yx}=\left<\max(\Re\{S_{yx}\}, 0_+)\right>_m$}%
\label{ssec:xsp-estimator-Neg-set-to-zero}
\begin{figure}[t]
\centering\namedgraphics{scale=0.8}{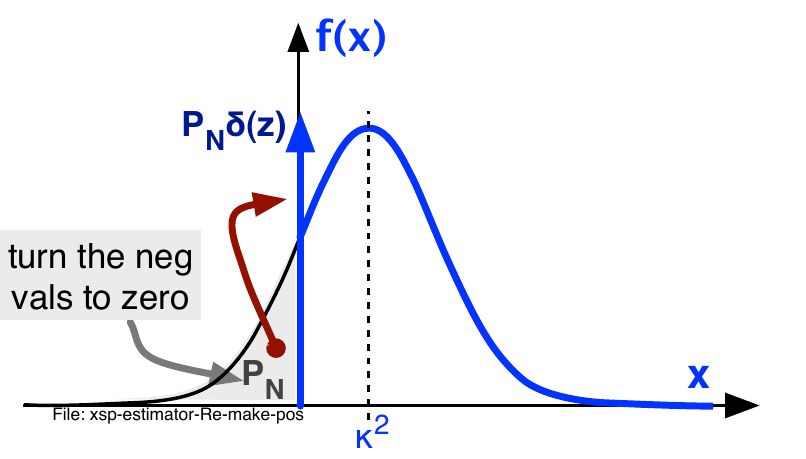}{\textwidth}
\caption{PDF of the estimator.}
\label{fig:xsp-estimator-Re-make-pos}
\end{figure}
Averaging $\Re\{S_{yx}\}$, the negative values are replaced with $0_+$.  The reason for using $0_+$ instead of just 0 is that $\lim_{x\rightarrow0_+}\log(x)$ exists, while $\lim_{x\rightarrow0}\log(x)$ does not.  The notation ``$0_+$'' is a nerdish replacement for the ``smallest positive floating-point number'' available in the computer.  This small number is equivalent to zero for all practical purposes, but never produces a floating-point error in the evaluation of the logarithm.
Since the negative values are replaced with zero, the PDF of this estimator (Fig.~\ref{fig:xsp-estimator-Re-make-pos}) derives from the PDF of $\Re\{\left<S_{yx}\right>_m\}$ replacing the negative-half-plane side with a Dirac delta function.

\subsection{Choice among the positive (biased) estimators}
\begin{figure}[t]
\centering\namedgraphics{scale=0.8}{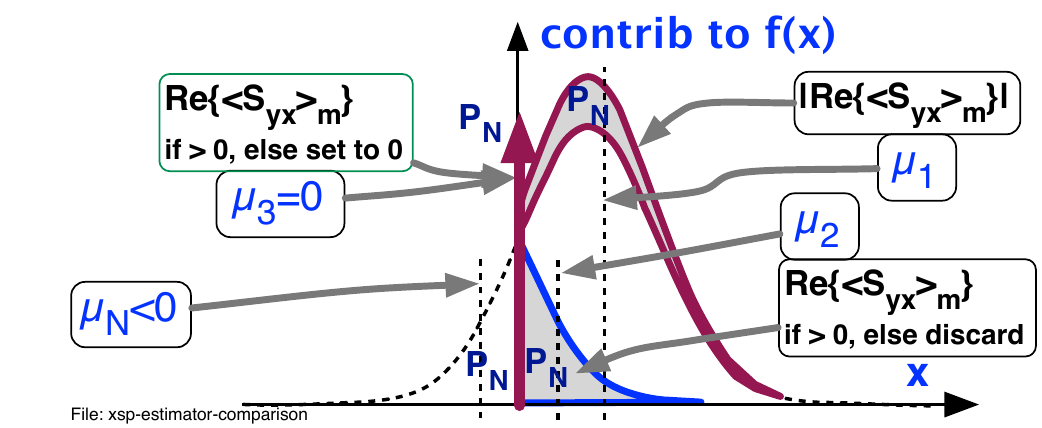}{\textwidth}
\caption{Comparison of the estimators based on $\Re\{S_{yx}\}$.}
\label{fig:xsp-estimator-comparison}
\end{figure}
Having accepted that an estimator suitable to logarithmic plot is positive, thus inevitably biased, the best choice is the estimator that exhibits the lowest variance and the lowest bias.
This criterion first excludes $|\left<S_{yx}\right>_m|$ in favor of one of the estimators based on $\Re\{\left<S_{yx}\right>_m\}$ because $\Im\{S_{yx}\}$ contains only the instrument background, which goes in both average (bias) and variance of $|\left<S_{yx}\right>_m|$.  Taking $\Im\{S_{yx}\}$ away, the estimator is necessarily based on $\Re\{\left<S_{yx}\right>_m\}$.

Then, we search for a suitable low-bias estimator with the heuristic reasoning shown in Figure \ref{fig:xsp-estimator-comparison}.

It is shown in Sec.~\ref{ssec:xsp-estimator-Real-part} that for large $m$ the PDF of $\Re\{\left<S_{yx}\right>_m\}$ is a Gaussian distribution with mean value $\kappa^2$ and variance $\sigma^2=\smash{\frac{1+2\kappa^2+2\kappa^4}{2m}}$.  
The probability of the events $\Re\{\left<S_{yx}\right>_m\}<0$ is represented in Fig.~\ref{fig:xsp-estimator-Re} as the grey area on the left-hand half-plane.  
These events have probability $P_N$.
Using the results of Section~\ref{ssec:xsp-gaussian}, the average of these negative events is
\begin{align*}
\mu_N=\int_{-\infty}^{\infty}x\,f_N(x)\:dx = \mu-\frac{1}{\frac12\text{erfc}\!\left(\frac{\mu}{\sqrt{2}\:\sigma}\right)} \: \frac{\sigma}{\sqrt{2\pi\exp(\mu^2/\sigma^2)}}
\qquad\text{(Eq.~\req{eqn:xsp-Gauss-mu-N})}~.
\end{align*}
The estimator is made positive by moving the area $P_N$ from the left-hand half-plane to the right-hand half-plane.  The bias depends on the shape taken by this area, and ultimately on the average associated to this shifted $P_N$.
By inspection on Fig.~\ref{fig:xsp-estimator-comparison} we notice that
\begin{description}
\item[Section \ref{ssec:xsp-estimator-Neg-discarded}.]  $\smash{\hat{S}_{yx}}=\Re\{\left<S_{yx}\right>_{m'}\}$ makes use only of the positive values, the negative values are discarded.
The PSD area associated to $P_N$ has the same shape of the right-hand side of the PSD\@.  We denote the average of this shape with $\mu_1$.

\item[Section \ref{ssec:xsp-estimator-Abs-Real-part}.]
$\smash{\hat{S}_{yx}}=|\Re\{\left<S_{yx}\right>_m\}|$.
The shadowed area associated to $P_N$ is flipped from the negative half-plane to the positive half-plane.  The average is $\mu_2=-\mu_N$.

\item[Section \ref{ssec:xsp-estimator-Neg-set-to-zero}.] $\smash{\hat{S}_{yx}}=\Re\{\left<\max(S_{yx}, 0_+)\right>_m\}$.
The shadowed area associated to $P_N$ collapses into a Dirac delta function.  The average is $\mu_3=0$.
\end{description}
From the graphical construction of Fig.~\ref{fig:xsp-estimator-comparison}, it is evident that \begin{align*}\mu_1>\mu_2>\mu_3~.\end{align*}
The obvious conclusion is that the preferred estimator is
\begin{align*}
\hat{S}_{yx}=\Re\left\{\left<\max(S_{yx}, 0_+)\right>_m\right\}
\qquad\text{(Preferred, Sec.~\ref{ssec:xsp-estimator-Neg-set-to-zero})}~.
\end{align*}
It is worth pointing out that the naif approach of just \emph{discarding the negative values} before averaging (Sec.~\ref{ssec:xsp-estimator-Neg-discarded}) turns out to be the \emph{worst choice} among the estimators we analyzed.

\subsection{The use of \boldmath$\Im\{\left<S_{yx}\right>_m\}$}
It has been shown in Sec.~\ref{sec:xsp-noise-rejection} (Eq.~\req{eqn:xsp-correl-dut-meas}) that all the DUT signal goes into $\Re\{S_{yx}\}$, and that $\Re\{S_{yx}\}$ contains only the instrument background.  More precisely, \req{eqn:xsp-correl-dut-meas} is rewritten as 
\begin{align*}
&S_{yx}=\tfrac1T\,\mathbb{E}\left\{\mathscr{A}+\imath\mathscr{B}+\mathscr{C}\right\}
\qquad\qquad\quad\qquad\text{(Eq.~\req{eqn:xsp-correl-dut-meas-ABC})}\\[1ex]
&\Re\{S_{yx}\}=\tfrac1T\,\mathbb{E}\left\{\mathscr{A}+\mathscr{C}\right\}
\quad\text{and}\quad
\Im\{S_{yx}\}=\tfrac1T\,\mathbb{E}\left\{\mathscr{B}\right\}
\end{align*}
where $\mathscr{A}$ and $\mathscr{B}$ come from the background have equal statistics, and $\mathscr{C}$ comes from the DUT spectrum.  Therefore
\begin{itemize}
\item $\Im\{\left<S_{yx}\right>_m\}$ is a good estimator of the background
\item the contrast $\Re\{\left<S_{yx}\right>_m\}-\Im\{\left<S_{yx}\right>_m\}$ is a good indicator of the averaging convergence to $S_{cc}$.
\end{itemize}

\section{Statistical independence on the frequency axis}
\begin{figure}[t]
\centering\namedgraphics{scale=0.7}{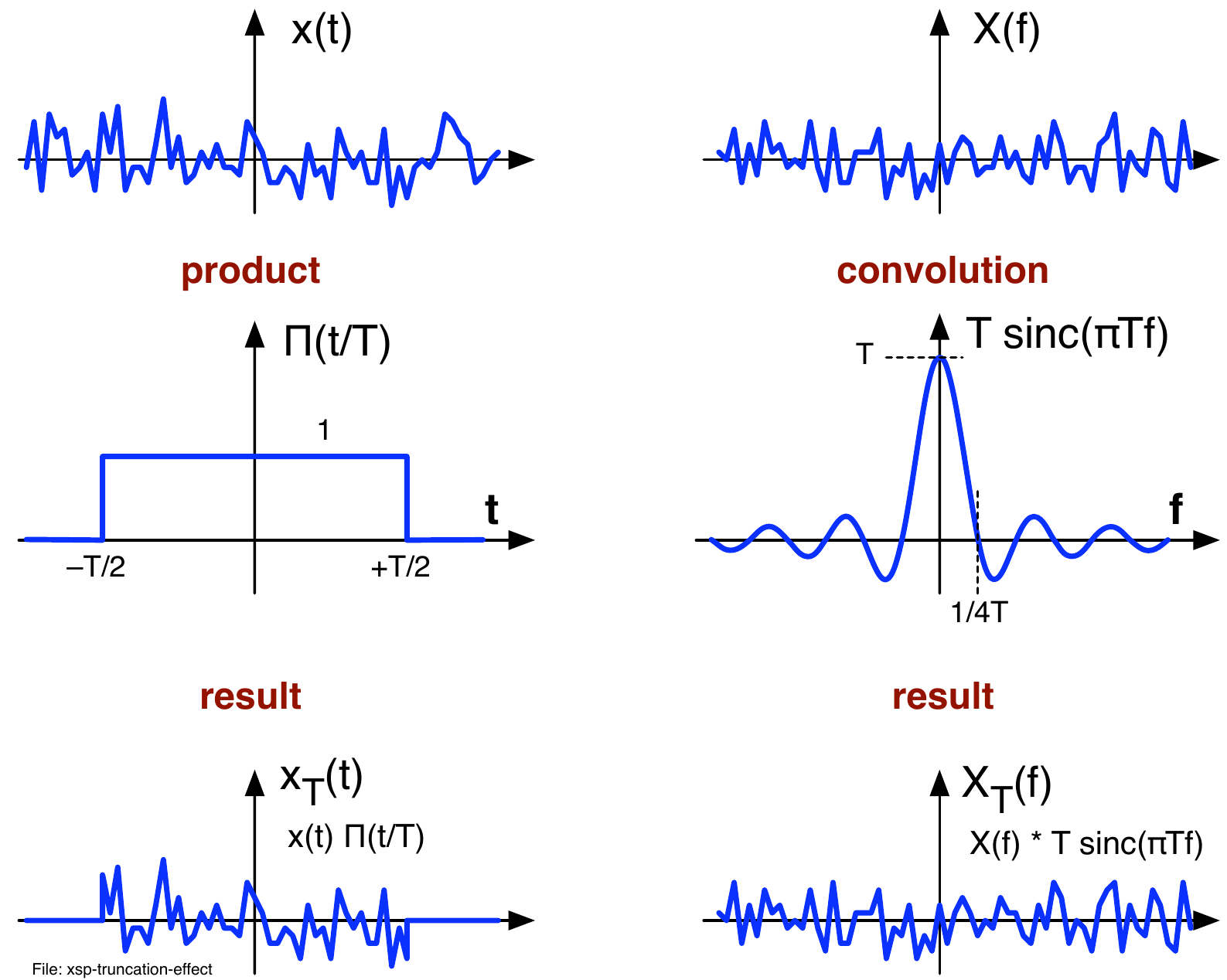}{\textwidth}
\caption{Effect of the finite duration of the measurement on the spectrum.}
\label{fig:xsp-truncation-effect}
\end{figure}

As a relevant property of white Gaussian noise, the Fourier transform is also Gaussian with all values on the frequency axis statistically-independent.  This property is taken as a good representation of the reality even in the case of discrete spectra measured on a finite measurement time $T$, and used extensively in this report.  
Yet, in a strictly mathematical sense time-domain truncation breaks the hypothesis of statistical independence in the frequency domain.
This happens because time truncation is equivalent to a multiplication by a rectangular pulse, which maps into a convolution by a sinc(\,) function in the frequency domain.  
This concept is shown in Fig.~\ref{fig:xsp-truncation-effect}, and expanded as follows
\begin{align*}
x(t) &\qquad\Rightarrow&x_T(t)&=x(t)\,\Pi(t/T)\\[1ex]
X(f) &\qquad\Rightarrow&X_T(f)&=x(t) \ast T\frac{\sin(\pi Tf)}{\pi Tf}
\end{align*}
where
\begin{align*}
\Pi(t)=\begin{cases}1&-1/2<t<1/2\\0&\text{elsewhere}\end{cases}
\qquad\leftrightarrow\qquad
\text{sinc}(f)=\frac{\sin(\pi f)}{\pi f}~.
\end{align*}
The consequences are the following.
\begin{itemize}
\item The side-lobes of $T$sinc$(Tf)$ cause energy leakage, thus a small correlation on the frequency axis.
\item Accuracy is reduced because each point collects energy from other frequencies.  This may show up in the presence of high peaks (50--60Hz, for example) or high roll-off bumps. 
\item One should question whether the number of degrees of freedom is reduced. 
\end{itemize}
The truncation function is called ``window'' on the front panel of analyzers, and sometimes ``taper'' in textbooks about spectral analysis.  
Reduced frequency leakage is obtained by a different choice of the truncation function, like the Bartlett (triangular), Hanning (cosine) or Parzen (cubic) window.

\section{Applications and experimental techniques}

\subsection{PM noise}\label{ssec:xsp-pm-noise}
The first application to frequency metrology was the measurement of Hydrogen masers \cite{vessot64nasa} in the early sixties.  Then, the method was used for the measurement of phase noise \cite{walls76fcs} in the seventies, but it found some popularity only in the nineties, when dual-channel FFT analyzers started to be available.  

\begin{figure}[t]
\centering\namedgraphics{scale=0.64, angle=0}{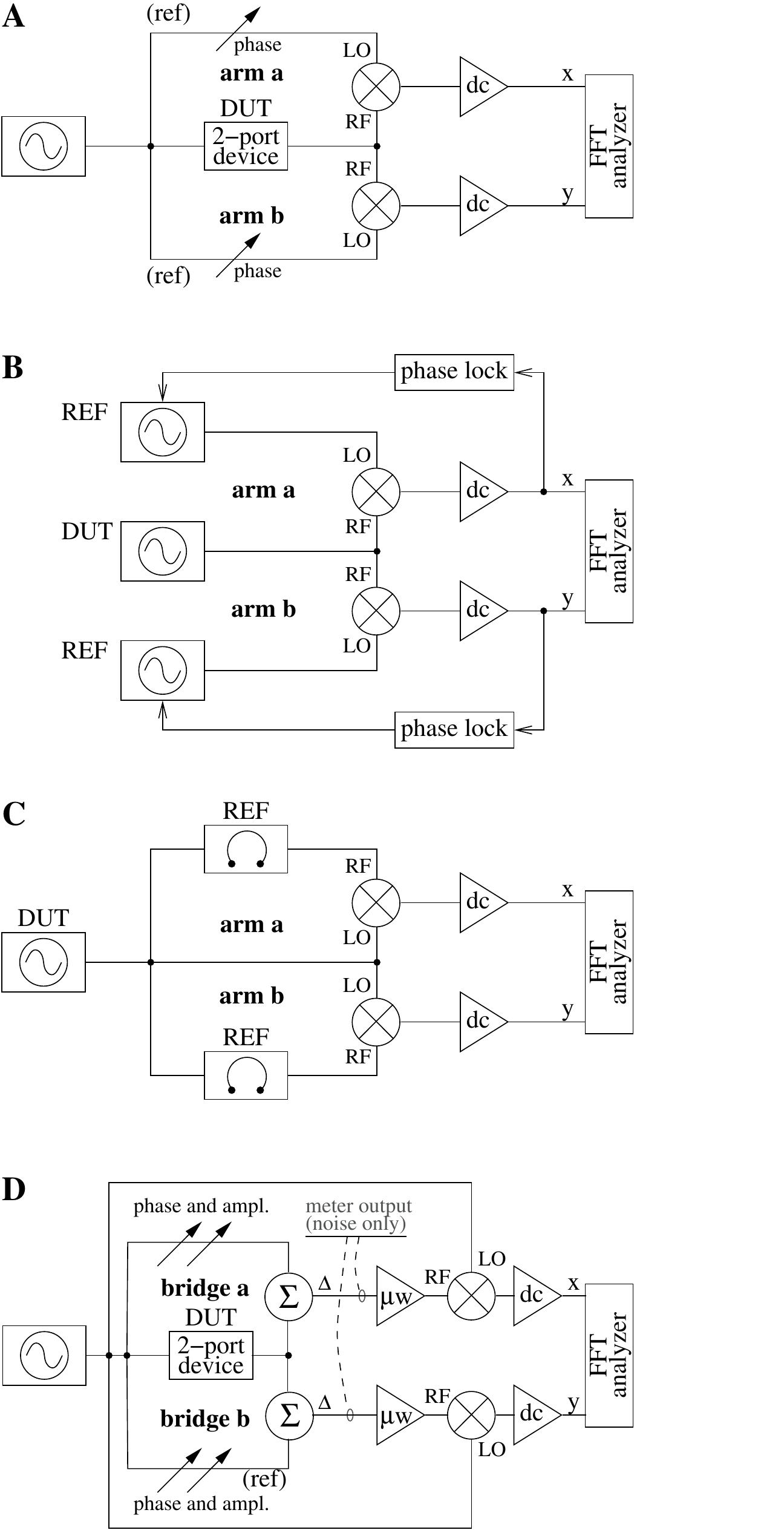}{\textwidth}
\caption{Basics schemes for the measurement of phase noise.}
\label{fig:xsp-sphi-schemes}
\end{figure}
Figure \ref{fig:xsp-sphi-schemes} shows some of the most popular schemes for the measurement of phase noise.  The mixer is a saturated phase-to-voltage converter in Fig.~\ref{fig:xsp-sphi-schemes} A-C, and a synchronous down-converter in Fig.~\ref{fig:xsp-sphi-schemes} D\@.  In all cases correlation is used to reject the noise of the two mixers.
The background noise turns out to be limited by the thermal homogeneity, instead of the absolute temperature referred to the carrier power.  This property was understood only after working on the scheme D \cite{rubiola2000rsi-correlation}.  At that time, the other schemes were already known.

The scheme A \cite{walls76fcs} is suitable to the measurement of low-noise two-port devices, mainly passive devices showing small group delay, so that the noise of the reference oscillator can be rejected.  

The scheme B consists of two separate PLLs that measure separately the oscillator under test.  Correlation rejects the noise of the two reference oscillators.  In this way, it is possible to measure an oscillator by comparing it to a pair of synthesizers, even if the noise of the synthesizers is higher than that of the oscillator.  This fact is relevant to the development of oscillator technology, when manufacturing makes it difficult to have the oscillator at the round frequency of the available standards, and also difficult to build two prototypes at the same frequency.

The scheme C derives from A after introducing a delay in the arms \cite{lance84}.  It can be implemented using either a pair of resonators or a pair a delay lines.  The use of the optical-fiber delay line is the most promising solution because the delay line can be adapted to the arbitrary frequency of the oscillator under test, while a resonator can not \cite{rubiola2005josab-delay-line}.  Correlation removes the fluctuations of the delay line \cite{salik04fcs-xhomodyne,salzenstein2007appa-dual-delay-line}.

The scheme D is based on a bridge that nulls the carrier before amplification and synchronous detection of the noise sidebands.  This scheme derives from the pioneering work of Sann \cite{sann68mtt}.  At that time, the mixer was used to down convert the fluctuation of the null at the output of a magic Tee.  Amplification of the noise sideband \cite{labaar82microw} and correlation \cite{rubiola2000rsi-correlation} were introduced afterwards.

With modern RF/microwave components, isolation between the two channels may not be a serious problem.  The hardware sensitivity is limited environmental effects, like temperature fluctuations and low-frequency magnetic fields, and by the AM noise.  The latter is taken in through the sensitivity of the mixer offset to the input power.  Only partial solutions are available \cite{rubiola2007uffc-am-to-pm-pollution}.

\subsection{AM noise}\label{ssec:xsp-am-noise}
\begin{figure*}[t]
\centering\textbf{A: amplitude noise of a RF/microwave source}\\[0.5em]
\namedgraphics{scale=0.8}{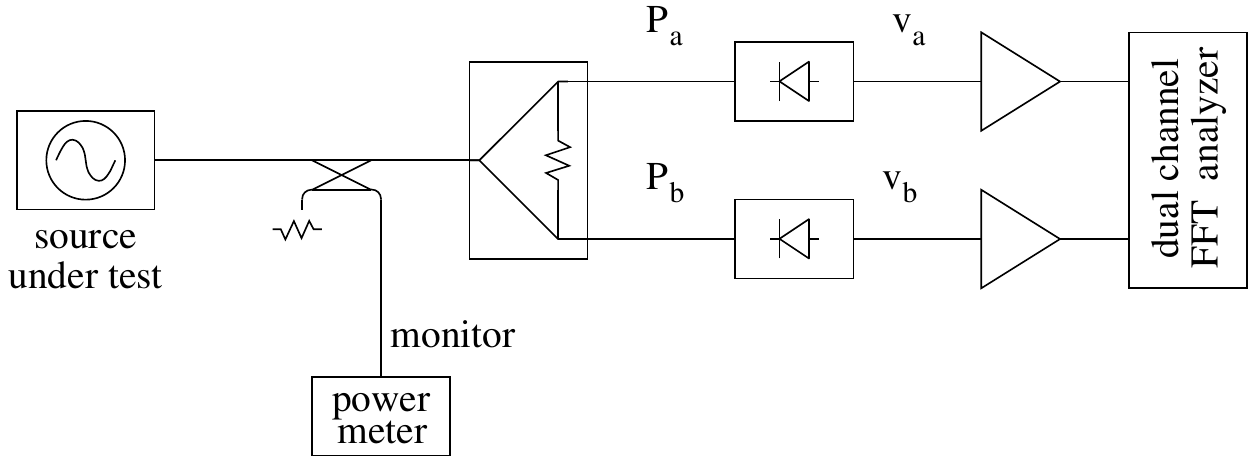}{\textwidth}\\[2em]
\centering\textbf{B: relative intensity noise (RIN) of a laser}\\[0.5em]
\centering\namedgraphics{scale=0.78}{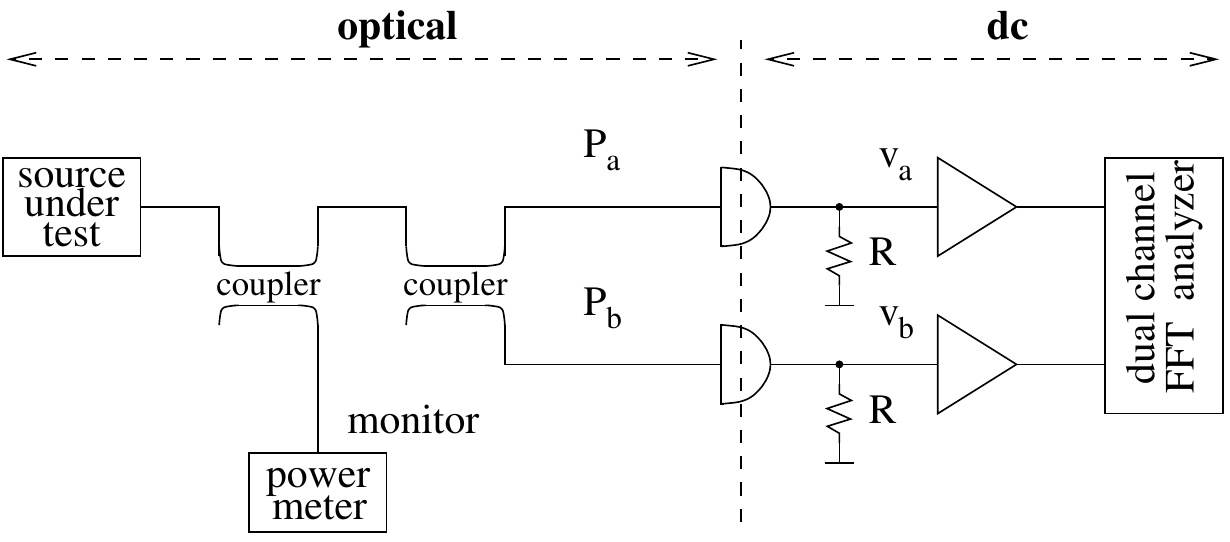}{\textwidth}\\[2em]
\centering\textbf{C: amplitude noise of a photonic RF/microwave source}\\[0.5em]
\centering\namedgraphics{scale=0.8}{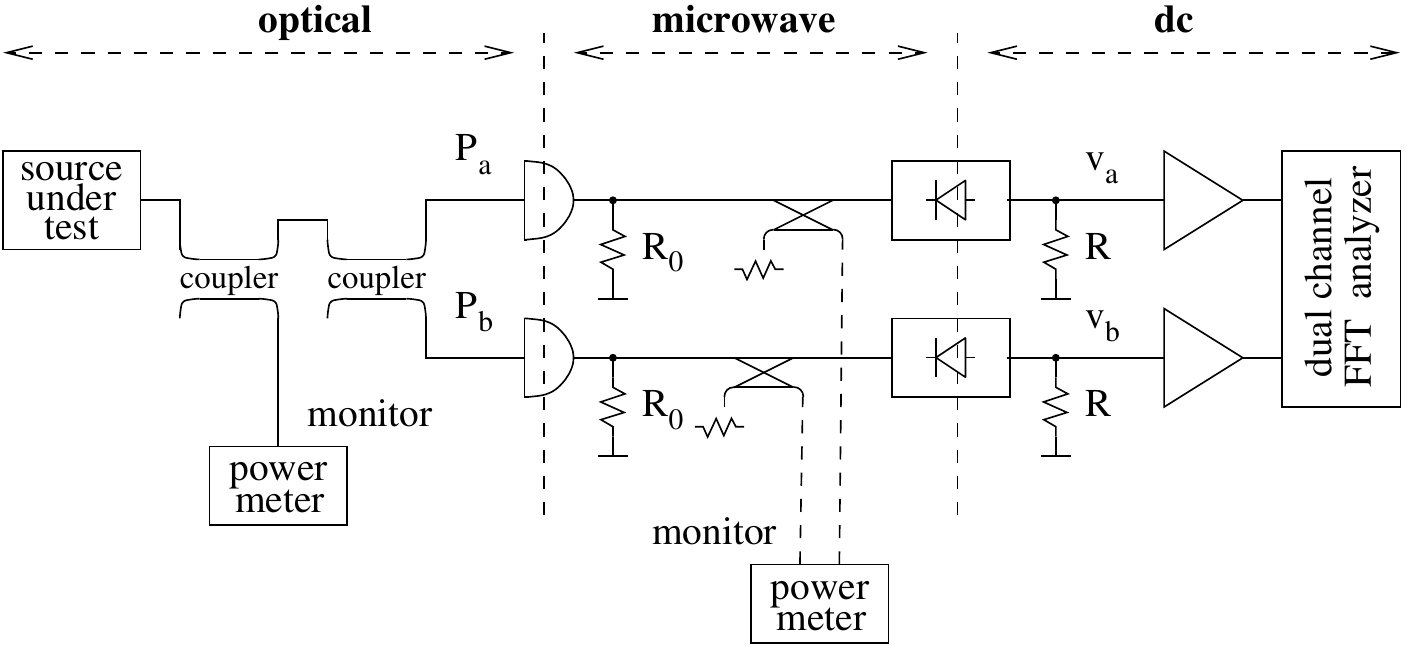}{\textwidth}
\caption{Basics schemes for the measurement of amplitude noise (from \cite{rubiola2005arxiv-am-noise}).}
\label{fig:mce-am}
\end{figure*}

Figure \ref{fig:mce-am} shows some schemes for the cross spectrum measurement of AM noise, taken from \cite{rubiola2005arxiv-am-noise}.

In Fig.~\ref{fig:mce-am}~A, two Schottky-diode or tunnel-diode passive power-detectors are used to measure simultaneously the power fluctuations of the source under test.  Isolation between channels is guaranteed by the isolation of the power splitter (18--20 dB) and by the fact that the power detectors do not send noise back to the input.
Correlation enables the rejection the single-channel noise.  

\begin{figure}[t]
\centering\namedgraphics{scale=0.63}{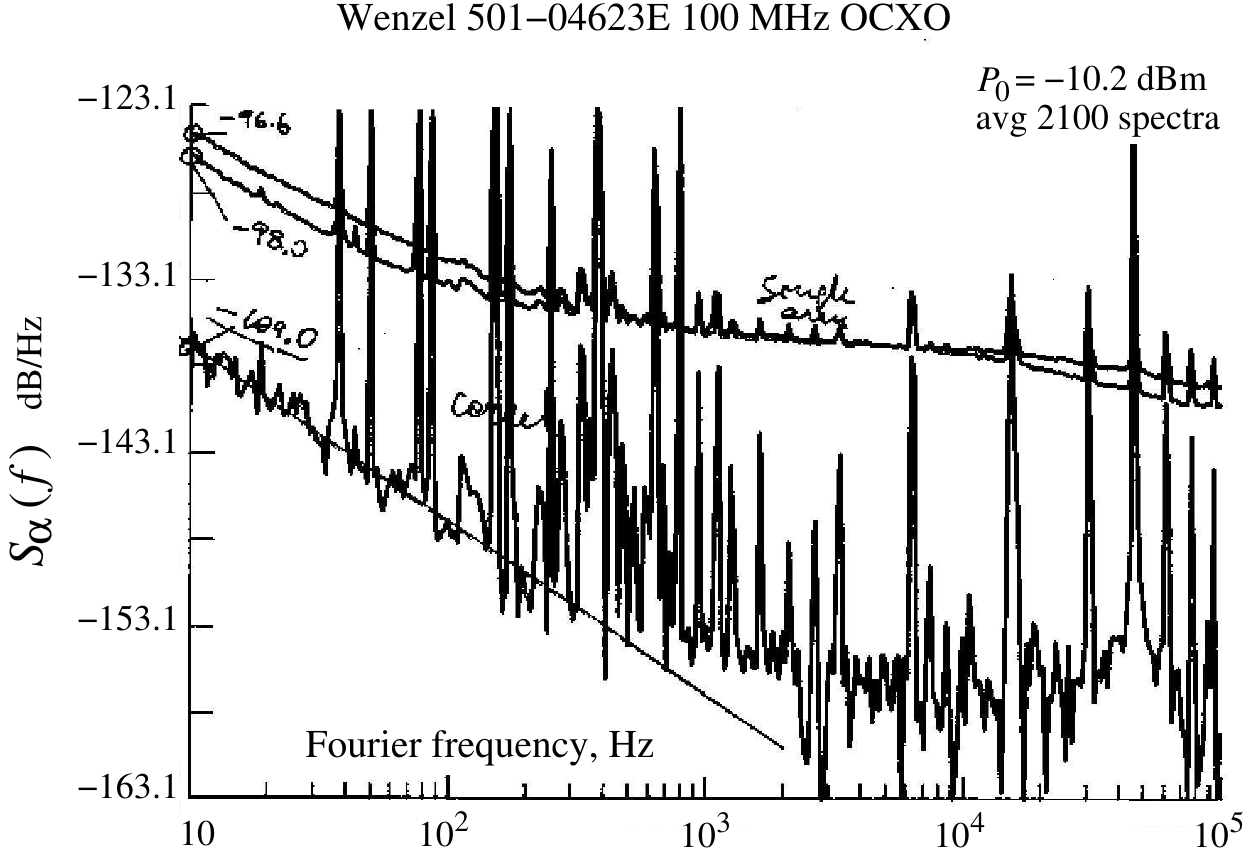}{\textwidth}
\caption{Example of cross spectrum measurement (amplitude noise of an oven-controlled quartz oscillator), taken from \cite{rubiola2005arxiv-am-noise}.}
\label{fig:am-wenzel-spectrum}
\end{figure}

As an example, Fig.~\ref{fig:am-wenzel-spectrum} shows the measurement of a quartz oscillator. Converting the $1/f$ noise into stability of the fractional amplitude $\alpha$, we get $\sigma_\alpha(\tau)=4.3{\times}10^{-7}$ (Allan deviation, constant vs.\ the measurement time $\tau$).  This oscillator exhibits the lowest AM noise measured in our laboratory.  The single-channel noise rejection achieved by correlation and averaging is more than 10 dB.

Figure \ref{fig:mce-am} B is the obvious adaptation of the scheme A to the measurement of the laser relative intensity noise (RIN).  We start using it routinely.  

The scheme of Fig.~\ref{fig:mce-am}~C, presently under study, is intended for the measurement of the microwave AM noise on the modulated light beam at the output of new generation of opto-electronic oscillators based on optical fibers \cite{Yao1996josab-oeo}, or based on whispering-gallery optical resonators. 

\begin{figure}
\centering\namedgraphics{scale=0.64}{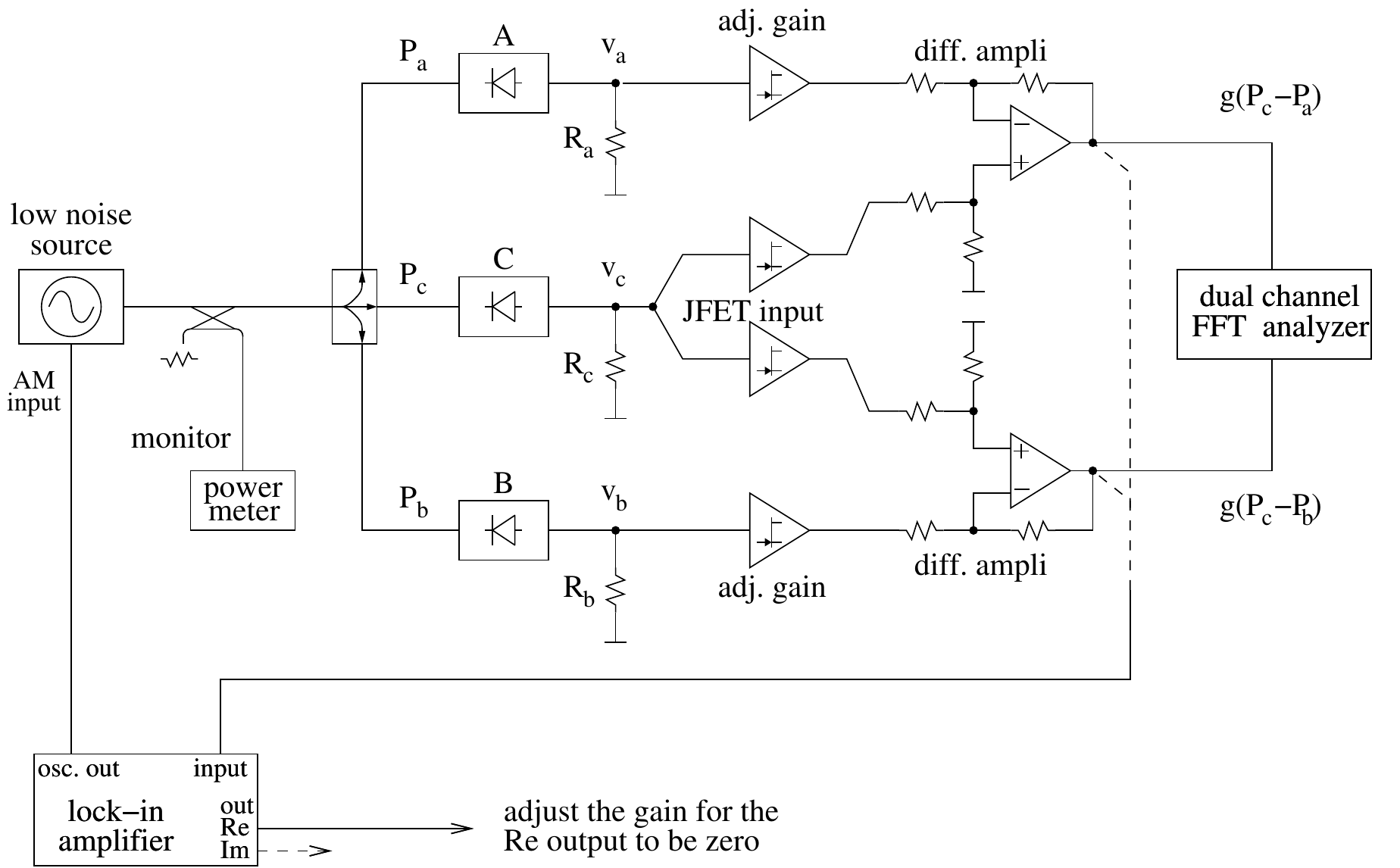}{\textwidth}
\caption{Measurement of the background noise of a power detector.}
\label{fig:xsp-am-detector-meas}
\end{figure}

\subsubsection{Single-chanel vs.\ dual-channel measurements}
In the measurement of PM noise it is more or less possible to test the  background of a single-channel instrument by removing the DUT\@.  This happens because we can always get the two phase-detector from a single oscillator, which is the phase reference.\footnote{This statement of course applies only to the background noise of the instrument.  When the instrument is used to measure an oscillator we need a reference oscillator, the noise of which must be validated separately.}
The correlation schemes are more complex than the single-channel counterparts, and sometimes difficult to operate.  Obviously, the experimentalist prefers the single-channel measurements and uses the correlation schemes only when the sensitivity of the former is insufficient.  

Conversely, the measurement of AM noise relies upon the power detector, which does not work without the source.  Thus we cannot remove the device under test, and of course we cannot asses the single-channel background noise \emph{of the instrument} in this way.  One can object that even in the case of PM noise we can not measure an oscillator in single-channel mode if we do not have a low-noise reference oscillator.  The difference is that in the case of PM noise we can at least validate the instrument, while in the case of AM noise we can not.

Another difference between AM and PM is that the phase detector is always more or less sensitive to AM noise \cite{rubiola2007uffc-am-to-pm-pollution}, while the amplitude detector is not sensitive to phase noise.  In correlation systems, this fact makes the channel separation simple to achieve and to test.

The conclusion is that the cross-spectrum measurement is inherently simpler with AM noise than with PM noise.

\subsection{Other applications}\label{ssec:xsp-other-applications}
Tracking back through the literature, the first use of the cross-spectrum was for the determination of the angular size of stellar radio sources \cite{hanbury-brown52nat}.  
In the case of a signal coming through two antennas separated by an appropriate baseline, the latter introduces a delay depending on the source direction in space.  Hence the useful signal $S_{cc}$ cannot be real.  Instead, the angle $\arctan{\Im/\Re}$ gives information on the source direction.  The very-large-baseline interferometry (VLBI) can be seen as a generalization of this method.

When the same method was applied to the intensity interferometer \cite{HanburyBrown1956nature-Correlation,HanburyBrown1956nature-Syrius}, an anti-correlation effect was discovered, due to the discrete nature of light.  This phenomenon, known as Hanbury Brown -- Twiss effect (HBT effect), was later observed also in microwave signals in photonic regime \cite{Gabelli2004prl-056801}, i.e., with $h\nu>kT$. 

The correlation method finds another obvious application in radiometry \cite{allred62jrnbs}, and of course in  Johnson thermometry, which is often considered a branch of radiometry.

Since the cross-spectrum enables to compare the PSD of two noise sources, it can be used to measure a temperature by comparing thermal noise to a reference shot noise.  The latter is in turn measured as a dc value by exploiting the property of Poisson processes that the variance can be calculated from the average.   
In a Tunnel junction, theory predicts the amount of shot and thermal noise.   This fact can be exploited for precision thermometry \cite{Spietz2003Science}, and ultimately to redefine the temperature in terms of fundamental constants.

The measurement of the low $1/f$ voltage fluctuations is an important diagnostic tool in semiconductor technology.  The field-effect transistors are suitable to this task because of the low bias current at the input.  In fact, the bias current flowing into the sample turns into a fully correlated voltage through the Ohm law.  Additionally, the electrode capacitance may limit the instrument sensitivity.  The reader can refer to \cite{sampietro99rsi} for a detailed treatise.

In metallurgy, the cross spectrum method has been used for the measurement of electromigration in thin metal films through the $1/f$ fluctuation of the conductor resistance.  This is relevant in microprocessor technology because the high current density in metal connexions can limit the life of the component and make it unreliable.  For this reason, Aluminum is no longer used.  
The high sensitivity is based on the idea that with white Gaussian noise $X'$ and $X''$ (real and imaginary part) are statistically independent.  Synchronously detecting the signal with two orthogonal references, it is therefore possible to reject the amplifier noise even if a single amplifier is shared by the two channel \cite{verbruggen89apa}.  
Adapting this idea to RF and microwaves is straightforward \cite{rubiola2002rsi-matrix}.  Unfortunately, we still have no application for this.

\appendix
\section{Mathematical background}

\subsection{Random variables and density functions}\label{ssec:xsp-rvar}
Let $\mathbf{x}$ a random variable and $x$ a variable.  Denoting with $\mathbb{P}\{\mathbf{e}\}$ the probability of the event $\mathbf{e}$, two relevant probability functions are associated with $\mathbf{x}$ and $x$, namely the cumulative density function $F(x)$ and the probability density function $f(x)$.  They are defined as
\begin{align}
&F(x) = \mathbb{P}\{\mathbf{x}<x\}&&\text{(cumulative density function, or CDF)}\\[1ex]
&f(x)\,dx = \mathbb{P}\{x<\mathbf{x}<x+dx\}&&\text{(probability density function, or PDF)}~.
\intertext{CDF and PDF are related by}
&F(x) = \int_{-\infty}^x f(x') dx'~.
\end{align}
The probability that $\mathbf{x}$ is in the interval $[a,b]$ is
\begin{align}
\mathbb{P}\{a<\mathbf{x}<b\} = F(b)-F(a) = \int_a^b f(x)\,dx~.
\end{align}
The probability that $\mathbf{x}$ takes any value is equal to one, thus 
\begin{align}
F(\infty)=1 \qquad\text{and}\qquad \int_{-\infty}^\infty f(x)\,dx = 1~.
\end{align}
The average and the variance of the random variable $\mathbf{x}$ are 
\begin{align}
\mathbb{E}\bigl\{\mathbf{x}\bigr\} &= \int_{-\infty}^\infty x\,f(x)\,dx &&\mathbb{E}\{\mathbf{x}\},~\text{(average)}\\
\mathbb{V}\bigl\{\mathbf{x}\bigr\} 
&=\mathbb{E}\bigl\{|\mathbf{x}-\mathbb{E}\{\mathbf{x}\}|^2\bigr\} 
	=\int_{-\infty}^\infty \bigl(x-\mathbb{E}\{\mathbf{x}\}\bigr)^2 \,f(x)\,dx
&&\text{(variance)}
\end{align}

\begin{figure}[t]
\centering\namedgraphics{scale=0.64}{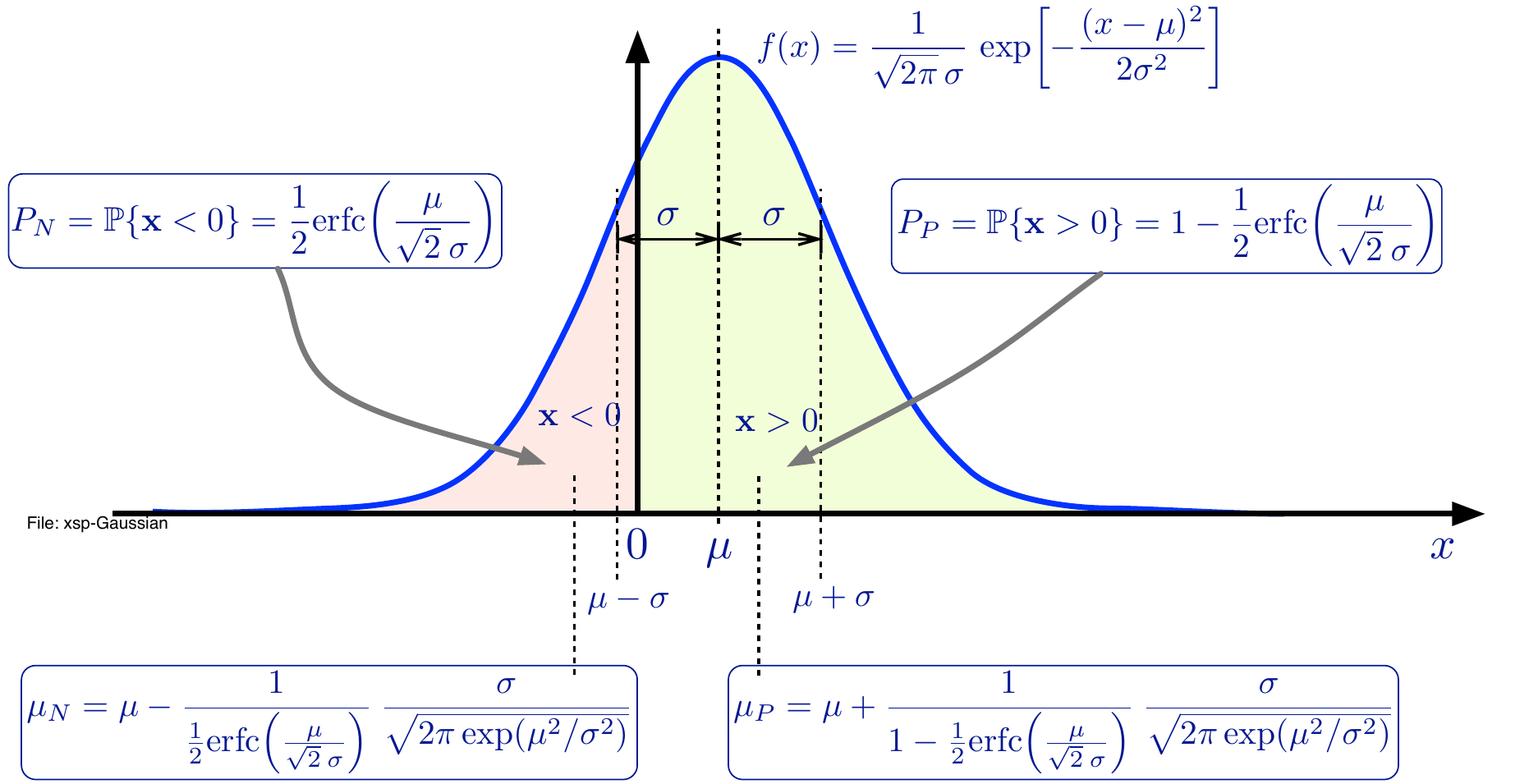}{\textwidth}
\caption{Gaussian (normal) PDF.}
\label{fig:xsp-Gaussian}
\end{figure}

\subsection{Gaussian (normal) distribution (Fig.~\ref{fig:xsp-Gaussian})}%
\label{ssec:xsp-gaussian}

The Gaussian (normal) distribution has the following main properties
\begin{align}
&f(x)=\frac{1}{\sqrt{2\pi}\,\sigma}\,\exp\left[-\frac{(x-\mu)^2}{2\sigma^2}\right]&&\text{(Gaussian PDF)}\\
&\mathbb{E}\{\mathbf{x}\}=\mu&&\text{(average)}\\[1ex] 
&\mathbb{V}\{\mathbf{x}\}=\sigma^2&&\text{(variance)}\\  
&P_N=\frac12 \text{erfc}\!\left(\frac{\mu}{\sqrt{2}\:\sigma}\right)
	&&(\mathbb{P}\{\mathbf{x}<0\})~.
	\label{eqn:xsp-Gauss-P-N}\\
&P_P=1-\frac12 \text{erfc}\!\left(\frac{\mu}{\sqrt{2}\:\sigma}\right)
	&&(\mathbb{P}\{\mathbf{x}>0\})
	\label{eqn:xsp-Gauss-P-P}
\end{align}

A new PDF is associated to the positive events
\begin{align}
&f_P(x)=\frac{1}{P_P} f(x) \,\mathfrak{u}(x)\\[1ex]
&\mu_P=\int_{-\infty}^{\infty}x\,f_P(x)\:dx = \mu + \frac{1}{1-\frac12\text{erfc}\!\left(\frac{\mu}{\sqrt{2}\:\sigma}\right)} \: \frac{\sigma}{\sqrt{2\pi\exp(\mu^2/\sigma^2)}}~.
\label{eqn:xsp-Gauss-mu-P}
\end{align}

Similarly, another PDF is associated to the negative events
\begin{align}
&f_N(x)=\frac{1}{P_N} f(x) \,\mathfrak{u}(-x)\\[1ex]
&\mu_N=\int_{-\infty}^{\infty}x\,f_N(x)\:dx = \mu-\frac{1}{\frac12\text{erfc}\!\left(\frac{\mu}{\sqrt{2}\:\sigma}\right)} \: \frac{\sigma}{\sqrt{2\pi\exp(\mu^2/\sigma^2)}}~.
\label{eqn:xsp-Gauss-mu-N}
\end{align}

The following integrals related to the Gaussian PDF are useful
\begin{align}
&\int_{-\infty}^{0}f(x)\:dx=\frac12\text{erfc}\!\left(\frac{\mu}{\sqrt{2}\:\sigma}\right)\\
&\int_{0}^{\infty}f(x)\:dx=1-\frac12\text{erfc}\!\left(\frac{\mu}{\sqrt{2}\:\sigma}\right)\\
&\int_{-\infty}^{0}x\,f(x)\:dx=\mu\,\frac12\text{erfc}\!\left(\frac{\mu}{\sqrt{2}\:\sigma}\right)
	-\frac{\sigma}{\sqrt{2\pi\exp(\mu^2/\sigma^2)}}\\
&\int_{0}^{\infty}x\,f(x)\:dx=\mu\left[1-\frac12\text{erfc}\!\left(\frac{\mu}{\sqrt{2}\:\sigma}\right)\right] +\frac{\sigma}{\sqrt{2\pi\exp(\mu^2/\sigma^2)}}~.
\end{align}

\subsubsection{Sum of zero-mean Gaussian variables}
Let $\mathbf{x}_1(t)$ and $\mathbf{x}_2(t)$ two random functions with Gaussian distribution, zero mean and variance $\sigma_1^2$ and $\sigma_2^2$.  
The sum $\mathbf{x}(t)=\mathbf{x}_1(t)+\mathbf{x}_2(t)$ is a random function with Gaussian distribution, zero mean and variance $\sigma^2=\sigma_1^2+\sigma_2^2$.  

\subsubsection{Sum of a nonzero-mean and a zero-mean Gaussian variable}
Let $\mathbf{x}_1(t)$ and $\mathbf{x}_2(t)$ two random functions with Gaussian distribution, and mean and variance $\mu_1\ne0$, $\sigma_1^2$, $\mu_2=0$, and $\sigma_2^2$.  The sum $\mathbf{x}(t)=\mathbf{x}_1(t)+\mathbf{x}_2(t)$ is a random function with Gaussian distribution, mean $\mu=\mu_1$ and variance $\sigma^2=\sigma_1^2+\sigma_2^2$.  

\subsubsection{Product of zero-mean Gaussian variables}
Let $\mathbf{x}_1(t)$ and $\mathbf{x}_2(t)$ two random functions with Gaussian distribution, zero mean and variance $\sigma_1^2$ and $\sigma_2^2$.  
The product $\mathbf{x}=\mathbf{x}_1(t)\,\mathbf{x}_2(t)$ is a random function with gaussian distribution, zero mean and variance $\sigma^2=\sigma_1^2\,\sigma_2^2$.  

\subsubsection{Fourier transform of a Gaussian variable}
Let $\mathbf{x}(t)$ a random process with Gaussian distribution and white spectrum, and $x(t)$ a realization.  The Fourier transform $X(f)=X'(f)+\imath X''(f)$ is a random process with white spectrum and zero-mean Gaussian distribution. This means that
\begin{enumerate}
\item At any frequency, the real part $X'(f)$ and the imaginary part $X''(f)$ are random variables statistically independent with equal variance.  
\item Given two frequencies $f_1$ and $f_2$ (or two separate frequency intervals), $X(f_1)$ and $X(f_2)$ are statistically independent.
\end{enumerate}
Interestingly, 
$|X|=\sqrt{(X')^2+(X'')^2}$
has Rayleigh distribution, and
$|X|^2=(X')^2+(X'')^2$
has $\chi^2$ distribution with 2 degrees of freedom.

\subsubsection{Discrete zero-mean Gaussian-distributed white noise}
It is often convenient to use the discrete Fourier transform and spectra.  Thus we refer to
\begin{align*}
X(f)~~~&\Rightarrow~~~X_{ij}=X'_{ij}+\imath X''_{ij}\\
S(f)~~~&\Rightarrow~~~S_{ij}=\frac1T\left(X_{ij}'^{\,2} + X_{ij}''^{\,2}\right)
\end{align*}
where the subscript $i$ denotes the $i$-th realization and the subscript $j$ denotes the discrete frequency.  The following properties hold for zero-mean white noise with Gaussian distribution.
\begin{enumerate}
\item $X_{ij}$ is zero-mean Gausian distributed.  Thus $X'_{ij}$ and $X''_{ij}$ are zero-mean Gaussian processes.
\item Different frequency.
	\vspace*{-\topsep}
	\begin{itemize} \setlength{\itemsep}{-\parsep}
	\item $X_{ij}$ and $X_{ik}$, $j\ne k$, are statistically independent.
	\item $\mathbb{V}\{X_{ij}\}=\mathbb{V}\{X_{ik}\}$ (energy equipartition).
	\end{itemize}
\item Real and imaginary part. 
	\vspace*{-\topsep}
	\begin{itemize} \setlength{\itemsep}{-\parsep}
	\item $X'_{ij}$ and $X''_{ij}$ are statistically independent.
	\item $\mathbb{E}\{X'_{ij}\}= 0$, and $\mathbb{E}\{X''_{ij}\}=0$ (zero mean).
	\item $\mathbb{V}\{X'_{ij}\}=\mathbb{V}\{X''_{ij}\}=\frac12\mathbb{V}\{X_{ij}\}$ (energy equipartition).
	\end{itemize}
\item Absolute square value $|X_{ij}|^2=|X'_{ij}|^2+|X''_{ij}|^2$.  Letting $\mathbb{V}\{X_{ij}\}=\sigma^2$,
	\vspace*{-\topsep}
	\begin{itemize} \setlength{\itemsep}{-\parsep}
	\item  $|X_{ij}|^2$ has $\chi^2$ distribution with two degrees of freedom.
	\item $\mathbb{E}\{|X_{ij}|^2\}=\sigma^2$ (average).
	\item $\mathbb{V}\{|X_{ij}|^2\}=\sigma^4$ (variance).
	\end{itemize}
\item Sum of two independent processes, $\mathbf{y}=\mathbf{x}_1+\mathbf{x}_2\leftrightarrow Y=X_1+X_2$. 
	\vspace*{-\topsep}
	\begin{itemize} \setlength{\itemsep}{-\parsep}
	\item $Y_{ij}$ is Gaussian distributed
	\item $\mathbb{V}\{Y_{ij}\}=\mathbb{V}\{X_{1\,ij}\}+\mathbb{V}\{X_{2\,ij}\}$.
	\end{itemize}
\item Product of two independent processes, $\mathbf{y}=\mathbf{x}_1 \mathbf{x}_2\leftrightarrow Y=X_1 * X_2$. 
	\vspace*{-\topsep}
	\begin{itemize} \setlength{\itemsep}{-\parsep}
	\item $Y_{ij}$ is Gaussian distributed
	\item $\mathbb{V}\{Y_{ij}\}=\mathbb{V}\{X_{1\,ij}\}+\mathbb{V}\{X_{2\,ij}\}$.
	\end{itemize}
\end{enumerate}

\subsection{Chi-square distribution (Table~\ref{tab:xsp-chi-square-prop})}\label{ssec:xsp-chisquare}
Let $\mathbf{x}_1$, $\mathbf{x}_2$, \ldots\ $\mathbf{x}_\nu$ a set of normal-distributed random variables with zero mean and variance equal one, and 
\begin{align}
\chi^2 = \sum_{i=1}^{\nu} \mathbf{x}_i^2
\end{align}
a new function called `chi-square' distribution with $\nu$ degrees of freedom.  The probability functions associated to $\mathbf{x}=\chi^2$ and the relevant parameters are
\begin{align}
&f(x)=\frac{\displaystyle x^{\frac12\nu-1} \, e^{-\frac12x}} {\Gamma\bigl(\frac{1}{2}\nu\bigr)\,2^{\frac12\nu }}\quad x\ge0
&&\text{(chi-square PDF)}\\[1ex]
&F(x)=1-\frac{\Gamma\bigl(\frac{1}{2}\nu, \frac12x\bigr)}{\Gamma\bigl(\frac{1}{2}\nu\bigr)}
= \frac{\gamma\bigl(\frac{1}{2}\nu, \frac12x\bigr)}{\Gamma\bigl(\frac{1}{2}\nu\bigr)}
&&\text{(chi-square CDF)}\\[1ex]
&\mathbb{E}\{\mathbf{x}\}=\nu&&\text{(average)}\\[1ex]  
&\mathbb{E}\{\mathbf{x}^2\}=\nu(\nu+2)&&\text{(2nd moment)}\\[1ex]  
&\mathbb{E}\{|\mathbf{x}-\mathbb{E}\{\mathbf{x}\}|^2\}=2\nu&&\text{(variance)}~.  
\end{align}
It follows immediately from the definition of $\chi^2$ that  the sum of $n$ random variables with $\chi^2$ distribution and $\nu_j$ degrees of freedom is $\chi^2$ distributed
\begin{align*}
\chi^2 = \sum_{j=1}^{n} \chi_{j}^2~,\quad \nu=\sum_{j=1}^{n} \nu_j~.
\end{align*}

\begin{table*}[t]
\begin{sideways}
\begin{minipage}{0.9\textheight}
\caption{Some properties of the $\chi^2$ distribution.}%
\label{tab:xsp-chi-square-prop}
\vspace*{5pt}
\def\rulethickness{0pt}
\def\pba#1{\parbox{15ex}{\setlength{\baselineskip}{2.5ex}\centering#1}}
\def\pbd#1{\parbox{26ex}{\setlength{\baselineskip}{2.5ex}\centering#1}}
\def\pbe#1{\parbox{26ex}{\setlength{\baselineskip}{2.5ex}\centering#1}}
\def\pbx#1{\parbox{14ex}{\setlength{\baselineskip}{2.5ex}\centering#1}}
\centering
\begin{tabular}{|c|c|c|c|c|}\hline\hline
\rule[-3ex]{\rulethickness}{7ex}%
distribution 
	& $\chi^2$ 
	& scaled $\chi^2$ 
	& \pbd{scaled $\chi^2$,\\average of $m$ real $\mathbf{x}^2_i$}
	& \pbe{scaled $\chi^2$, average of $m$\\[0.5ex] complex $\mathbf{x}^2_i=\mathbf{x}_{i}'^2+\imath\mathbf{x}_{i}''^2$}\\\hline 
rule\rule[-4.5ex]{\rulethickness}{10ex}
	&$\displaystyle\mathbf{x}=\chi^2=\sum_{i=1}^{\nu}\mathbf{x}_i^2$
	&$\displaystyle\mathbf{x}=\chi^2=\sum_{i=1}^{\nu}\mathbf{x}_i^2$
	&$\displaystyle\mathbf{x}=\frac1m\sum_{i=1}^{\nu}\mathbf{x}_i^2$
	&$\displaystyle\mathbf{x}=\frac1m\sum_{i=1}^{\nu}\left(\mathbf{x}_{i}'^{\,2}+\mathbf{x}_{i}''^{\,2}\right)$\\
\rule[-1ex]{\rulethickness}{1ex}%
	&$\mathbb{E}\{\mathbf{x}_i\}=0$
	&$\mathbb{E}\{\mathbf{x}_i\}=0$
	&$\mathbb{E}\{\mathbf{x}_i\}=0$
	&$\mathbb{E}\{\mathbf{x}'_i\}=\mathbb{E}\{\mathbf{x}''_i\}=0$\\
\rule[-2ex]{\rulethickness}{4.5ex}%
	&$\mathbb{V}\{\mathbf{x}_i\}=1$
	&$\mathbb{V}\{\mathbf{x}_i\}=\sigma^2$
	&$\mathbb{V}\{\mathbf{x}_i\}=\sigma^2$
	&$\mathbb{V}\{\mathbf{x}'_i\}=\mathbb{V}\{\mathbf{x}''_i\}=\sigma^2/2$\\\hline
\rule[-3.5ex]{\rulethickness}{8ex}%
transformation
	&\pbx{none\\$f(x)=k_\nu(x)$}
	&$\displaystyle f(x)=\frac{1}{\sigma^2}\:k_\nu\Bigl(\frac{x}{\sigma^2}\Bigr)$
	&$\displaystyle f(x)=\frac{m}{\sigma^2}\:k_m\Bigl(\frac{mx}{\sigma^2}\Bigr)$
	&$\displaystyle f(x)=\frac{2m}{\sigma^2}\:k_{2m}\Bigl(\frac{2mx}{\sigma^2}\Bigr)$\\\hline
\rule[-4ex]{\rulethickness}{9.5ex}%
\pba{probability density function}
	&$\displaystyle f(x)=\frac{x^{\frac12\nu-1}\:e^{-\frac12x}}{\Gamma\bigl(\frac12\nu\bigr)\:2^{\frac12\nu}}$
	&$\displaystyle f(x)=\frac{x^{\frac12\nu-1}\:e^{-\frac{x}{2\sigma^2}}}{\sigma^\nu\:\Gamma\bigl(\frac12\nu\bigr)\:2^{\frac12\nu}}$
	&$\displaystyle f(x)=\frac{m^{\frac12m}\:x^{\frac12m-1}\:e^{-\frac{mx}{2\sigma^2}}}{\sigma^m\:\Gamma\bigl(\frac12m\bigr)\:2^{\frac12m}}$
	&$\displaystyle f(x)=\frac{m^m\:x^{\frac12m-1}\:e^{-\frac{mx}{\sigma^2}}}{\sigma^{2m}\:\Gamma(m)}$\\\hline
\rule[-2ex]{\rulethickness}{6ex}%
\pbx{average\\$\mathbb{E}\{\mathbf{x}\}$}
	&$\nu$
	&$\nu\sigma^2$
	&$\sigma^2$
	&$\sigma^2$\\\hline
\rule[-2.5ex]{\rulethickness}{6.5ex}%
\pba{variance\\$\mathbb{E}\{|\mathbf{x}-\mathbb{E}\{\mathbf{x}\}|^2\}$}
	&$2\nu$
	&$2\nu\sigma^4$
	&$2\sigma^4/m$
	& $\sigma^4/m$\\\hline\hline
\end{tabular}
\end{minipage}
\end{sideways}
\end{table*}

In the general case, the variance of $\mathbf{x}_1$ \ldots $\mathbf{x}_\nu$ is $\sigma^2\ne1$.  This is solved with the transformation $\mathbf{x}\rightarrow\mathbf{x}/\sigma^2$.  Thus $f(x)=\smash{\frac{1}{\sigma^2}}\left[f(x)\right]_\text{var=1}$, and\sidenote{Cram\'er p.\,236}
\begin{align}
&f(x)=\frac{\displaystyle x^{\frac12\nu-1} \, e^{-\frac12\frac{x}{\sigma^2}}} {\sigma^\nu\,\Gamma\bigl(\frac{1}{2}\nu\bigr)\,2^{\frac12\nu }}\quad x\ge0
&&\text{(chi-square PDF)}\\[1ex]
&\mathbb{E}\{\mathbf{x}\}=\sigma^2\nu&&\text{(average)}\\[1ex]  
&\mathbb{E}\{\mathbf{x}^2\}=\sigma^4\nu(\nu+2)&&\text{(2nd moment)}\\[1ex]  
&\mathbb{E}\{|\mathbf{x}-\mathbb{E}\{\mathbf{x}\}|^2\}=2\sigma^4\nu&&\text{(variance)}~.  
\end{align}

\subsection{Rayleigh distribution}\label{ssec:xsp-rayleigh}
Let $\mathbf{x}_1$ and $\mathbf{x}_2$, two independent random functions with Gaussian distribution, zero mean and equal variance $\sigma$, and 
\begin{align}
\mathbf{x}=\sqrt{\mathbf{x}_1^2+\mathbf{x}_2^2}
\label{eqn:xsp-rayleigh-y}
\end{align}
a new random function.  This function has Rayleigh probability density function\sidenote{Checked}
\begin{align}
\label{eqn:xsp-rayleigh-pdf}
&f(x)=\frac{x}{\sigma^2}\,\exp\left(-\frac{x^2}{2\sigma^2}\right),\quad y>0&&\text{(Rayleigh PDF)}\\
&\mathbb{E}\{\mathbf{x}\}=\sqrt{\frac{\pi}{2}}\:\sigma&&\text{(average)}\\[1ex]
&\mathbb{E}\{\mathbf{x}^2\}=2\sigma^2&&\text{(2nd moment)}\\
&\mathbb{V}\{\mathbf{x}\}=\mathbb{E}\{|\mathbf{x}-\mathbb{E}\{\mathbf{x}\}|^2\}=\frac{4-\pi}{2}\sigma^2&&\text{(variance)}~.
\end{align}
The functions $\mathbf{x}_1(t)$ and $\mathbf{x}_2(t)$ can be interpreted as the random amplitude of two orthogonal vectors, or the real and imaginary part of a complex random function.  Following this interpretation, $\mathbf{x}(t)$ is the absolute value of the vector sum. 
Table \ref{tab:xsp-rayleigh} reports some useful numerical values related to the $\sigma^2=1/2$ Rayleigh distribution.

A case of interest in averaged measurement is $\sigma^2=1/2m$, which yields
\begin{align}
\label{eqn:xsp-rayleigh-avg-m}
&\mathbb{E}\{\mathbf{x}\}=\sqrt{\frac{\pi}{4m}}=\frac{0.886}{\sqrt{m}}&&\text{(average)}\\[1ex]
\label{eqn:xsp-rayleigh-2nd-moment-m}
&\mathbb{E}\{\mathbf{x}^2\}=\frac{1}{m}&&\text{(2nd moment)}\\[1ex]
\label{eqn:xsp-rayleigh-var-m}
&\mathbb{V}\{\mathbf{x}\}=\mathbb{E}\{|\mathbf{x}-\mathbb{E}\{\mathbf{x}\}|^2\}=\left(1-\frac{\pi}{4}\right)\frac{1}{m}=\frac{0.215}{m}&&\text{(variance)}\\[1ex]
\label{eqn:xsp-rayleigh-dev-m}
&\sqrt{\mathbb{V}\{\mathbf{x}\}}=\sqrt{\frac1m\left(1-\frac{\pi}{4}\right)}=\frac{0.463}{\sqrt{m}}&&\text{(deviation)}\\[2ex]
\label{eqn:xsp-rayleigh-dev-avg-ratio}
&\frac{\sqrt{\mathbb{V}\{\mathbf{x}\}}}{\mathbb{E}\{\mathbf{x}\}}
=\sqrt{\frac{4}{\pi}-1}=0.523\quad\text{(independent of $m$)}&&\text{(dev/avg)}~.
\end{align}

\begin{table}[t]
\begin{center}
\caption{Relevant values for the Rayleigh distribution.}
\label{tab:xsp-rayleigh}
\vspace{1ex}
\def\vr{\rule[-2.5ex]{0pt}{6.8ex}}
\begin{tabular}{|c|c|}\hline
\multicolumn{2}{|c|}{\rule[-1.2ex]{0pt}{3.5ex}Rayleigh distribution with $\sigma^2=1/2$}\\\hline
\rule[0ex]{0pt}{2.2ex}quantity &value\\
\rule[-1.2ex]{0pt}{2.2ex}with $\sigma^2=1/2$ &[$10\log(~)$, dB]\\\hline
\vr$\text{average}=\displaystyle\sqrt{\frac\pi4}$ 
&\begin{tabular}{c}0.886\\$[-0.525]$\end{tabular}\\\hline
\vr$\text{deviation}=\displaystyle\sqrt{1-\frac\pi4}$ 
&\begin{tabular}{c}0.463\\$[-3.34]$\end{tabular}\\\hline
\vr$\unit{\displaystyle\frac{dev}{avg}}=\displaystyle\sqrt{\frac4\pi-1}$
&\begin{tabular}{c}0.523\\$[-2.8]$\end{tabular}\\\hline
\vr$\unit{\displaystyle\frac{avg+dev}{avg}}=\displaystyle1+\sqrt{\frac4\pi-1}$ 
&\begin{tabular}{c}1.523\\$[+1.83]$\end{tabular}\\\hline
\vr$\unit{\displaystyle\frac{avg-dev}{avg}}=\displaystyle1-\sqrt{\frac4\pi-1}$ 
&\begin{tabular}{c}0.477\\$[-3.21]$\end{tabular}\\\hline
\end{tabular}
\end{center}
\end{table}

\section{A short introduction to AM and PM noise}
Phase noise (PM noise) is a well established subject, clearly explained in numerous classical references, among which we prefer \cite{rutman78pieee,kimball:precise-frequency,ccir90rep580-3,ieee99std1139} and \cite[vol.\,1, chap.\,2]{vanier:frequency-standards}.  Amplitude noise (AM noise), far less studied than PM noise, is described in similar manner.  Refer to \cite{rubiola2005arxiv-am-noise} for a general introduction to AM noise.  Only a brief introduction to AM/PM noise is given here, aimed at recalling the vocabulary.

The quasi-perfect sinusoidal signal of frequency $\nu_0$, of random amplitude fluctuation $\alpha(t)$, and of random phase fluctuation $\varphi(t)$ is
\begin{align}
v(t)=\left[1+\alpha(t)\right]\cos\left[2\pi\nu_0t+\varphi(t)\right]~.
\end{align}
We may need that $|\alpha(t)|\ll1$ and that $|\varphi(t)|\ll1$ or $|\dot{\varphi}(t)|\ll1$ during the measurement.

\subsection{Spectral representation of PM noise}
Phase noise is generally reported in terms of the PSD (power spectral density) $S_\varphi(f)$. 
In experiments, the single-sided PSD $\smash{S^I_\varphi}(f)$ is preferred to the two-sided PSD $\smash{S^{I\!I}_\varphi}(f)$ because the negative frequencies are redundant for real signals.  Complex or imaginary signals do not exist in this context.  Thus, energy conservation requires that $S^I_\varphi(f)=2S^{I\!I}_\varphi(f)$ for $f>0$.  Since now, we use $S_{\varphi}(f)$ as the single-sided PSD, dropping the superscript `$I$.' 
  
\begin{figure}
\centering\namedgraphics{scale=0.8}{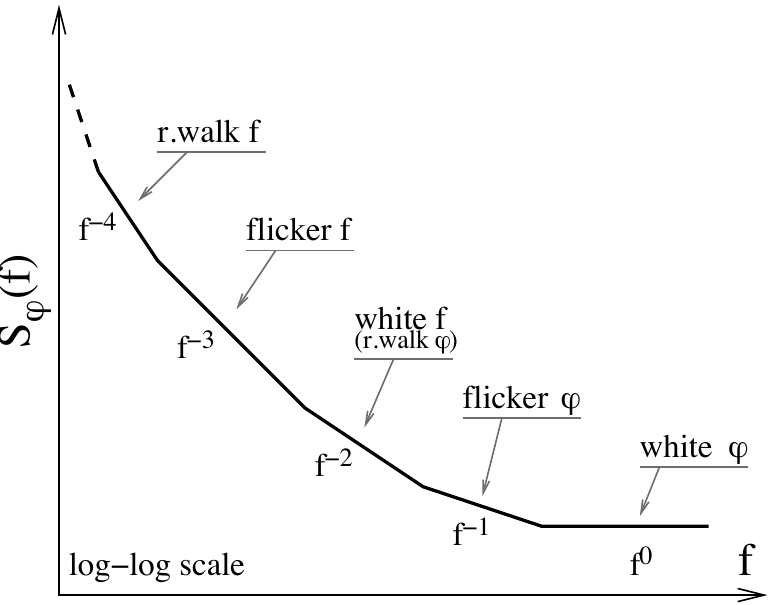}{\textwidth}
\caption{Power law model for $S_\varphi(f)$ (from \cite{rubiola2005josab-delay-line}).}
\label{fig:ddl-power-law}
\end{figure}
A model that has been found useful to describe accurately the phase noise of oscillator and components is the power law, shown in Fig.~\ref{fig:ddl-power-law}
\begin{align}
S_{\varphi}(f)&= \sum_{n=-4}^{0}b_if^n \qquad\text{(power law)}~.
\label{eqn:xsp-power-law-sphi}
\end{align}
This model relies on the fact that white ($f^0$) and flicker ($1/f$) noises exist per-se, and that phase integration ($\times1/f^2$) is present in oscillators.  If needed, the model can be extended to steeper processes, that is, $n<-4$.

When frequency noise (FM noise) is preferred to phase noise, the fractional frequency fluctuation $y(t)=\dot{\varphi}(t)/2\pi\nu_0$ is probably the most useful quantity.   Using the power law, the spectrum $S_y(f)$ is written as  
\begin{align}
S_{y}(f)=\frac{f^2}{\nu_0^2}\:S_{\varphi}(f) = \sum_{n=-2}^{2}h_if^n~.
\label{eqn:xsp-sy}
\end{align}

\subsection{Spectral representation of AM noise}

Amplitude noise is described in the same way of phase noise or frequency noise, and for the same reasons we use the power law
\begin{align}
S_{\alpha}(f) &= \sum_{n=-2}^{0}h_if^n \qquad\text{(power law)}~.
\label{eqn:xsp-power-law-salpha}
\end{align}
Yet, the set of processes found in practice is often limited to white and flicker noise, and to random walk.  Steeper processes ($n<-2$), when present, tend to be confined to a limited region of the spectrum.  They vanish at very low frequencies, otherwise the amplitude would diverge rapidly.
Notice that we use the coefficients $h_i$ as for FM noise instead of the $b_i$ used with PM noise.  The reason is that the formulae for the Allan variance (see below) are formally equal.

\subsection{Two-sample (Allan) variance}

Another tool often used is the Allan variance $\sigma^2_y(\tau)=\smash{\mathbb{E}\{|\overline{y}_{k+1}-\mathbb{E}\{\overline{y}_{k}\}|^2\}}$, where $\smash{\overline{y}_{k}}$ is the average of $y(t)$ over the $k$-th contiguous time slot of duration $\tau$, spanning from $k\tau$ to $(k+1)\tau$.  For the most useful frequency-noise processes, the relation between $\sigma^2_y(\tau)$ and $S_y(f)$ is
\begin{align}
\sigma^2_y(\tau)=\begin{cases} 
	\displaystyle
	\frac{h_0}{2\tau}	& \text{white frequency noise}\\[1ex]
	h_{-1}\,2\ln(2)		& \text{flicker of frequency}\\[1ex]
	\displaystyle
	h_{-2}\,\frac{(2\pi)^2}{6}\,\tau	& \text{random walk of frequency}\\[1ex]
	\ldots	& \text{(other phenomena, if any)}
	\end{cases}
\end{align}
Similarly, letting $\sigma^2_\alpha(\tau)=\smash{\mathbb{E}\{|\overline{\alpha}_{k+1}-\mathbb{E}\{\overline{\alpha}_{k}\}|^2\}}$, the AM-noise variance is
\begin{align}
\sigma^2_\alpha(\tau)=\frac{h_0}{2\tau}+h_{-1}\,2\ln(2)+h_{-2}\,\frac{(2\pi)^2}{6}\,\tau+\ldots
\end{align}

\subsection*{Acknowledgments}
ER owes gratitude to Charles Greenhall (NASA/Caltech Jet Propulsion Laboratory, USA) and to Michele Elia (Politecnico di Torino, Italy) for a wealth of discussions about statistics; to Lute Maleki (OEwaves, USA) for offering an important scientific opportunity, relevant to this work and for discussions; to Vincent Giordano (FEMTO-ST, France) for numerous discussion and for supporting me over more than ten years.

\def\bibfile#1{/Users/rubiola/Documents/Articles/Bibliography/#1}
\addcontentsline{toc}{section}{References}
\bibliographystyle{amsalpha}
\bibliography{\bibfile{ref-short},\bibfile{references},\bibfile{rubiola}}

\newcommand{\etalchar}[1]{$^{#1}$}
\providecommand{\bysame}{\leavevmode\hbox to3em{\hrulefill}\thinspace}
\providecommand{\MR}{\relax\ifhmode\unskip\space\fi MR }
\providecommand{\MRhref}[2]{%
  \href{http://www.ams.org/mathscinet-getitem?mr=#1}{#2}
}
\providecommand{\href}[2]{#2}
\begin{thebibliography}{HBJDG52}

\bibitem[All62]{allred62jrnbs}
C.~M. Allred, \emph{A precision noise spectral density comparator}, J. Res.\
  NBS \textbf{66C} (1962), 323--330.

\bibitem[{CCI}90]{ccir90rep580-3}
{CCIR Study Group VII}, \emph{Characterization of frequency and phase noise,
  {R}eport no.\ 580-3}, Standard Frequencies and Time Signals, Recommendations
  and Reports of the {CCIR}, vol. {VII} (annex), International
  Telecommunication Union {(ITU)}, Geneva, Switzerland, 1990, pp.~160--171.

\bibitem[Cra46]{Cramer:statistics}
Harald Cram\'{e}r, \emph{Mathematical methods of statistics}, Princeton, 1946.

\bibitem[DR58]{Davenport-Root:noise}
Wilbur~D. Davenport, Jr and William~L. Root, \emph{An introduction to random
  signals and noise}, Mc{G}raw Hill, New York, 1958, (Reprinted by the IEEE
  Press, New York, 1987).

\bibitem[Fell2]{Feller:probability}
William Feller, \emph{An introduction to probability theory and its
  applications}, 2nd ed., vol. 2 volumes, Wiley, New York, 1957 (vol.\,1), 1971
  (vol.\,2).

\bibitem[GRF{\etalchar{+}}04]{Gabelli2004prl-056801}
J.~Gabelli, L.-H. Reydellet, G.~F\`{e}ve, J.-M. Berroir, B.~Pla\c{c}ais,
  P.~Roche, and D.~C. Glattli, \emph{{H}anbury {B}rown -- {T}wiss correlations
  to probe the population statistics of {GHz} photons emitted by conductors},
  Phys.\ Rev.\ Lett. \textbf{93} (2004), no.~5, 056801.

\bibitem[HBJDG52]{hanbury-brown52nat}
R.~Hanbury~Brown, R.~C. Jennison, and M.~K. Das~Gupta, \emph{Apparent angular
  sizes of discrete radio sources}, Nature \textbf{170} (1952), no.~4338,
  1061--1063.

\bibitem[HBT56a]{HanburyBrown1956nature-Correlation}
R.~Hanbury~Brown and R.~Q. Twiss, \emph{Correlation between photons in two
  coherent beams of light}, Nature \textbf{177} (1956), 27--29.

\bibitem[HBT56b]{HanburyBrown1956nature-Syrius}
\bysame, \emph{A test of a new type of stellar interferometer on {S}irius},
  Nature \textbf{178} (1956), 1046--1048.

\bibitem[Kim97]{kimball:precise-frequency}
H.~G. Kimball (ed.), \emph{Handbook of selection and use of precise frequency
  and time systems}, ITU, 1997.

\bibitem[Lab82]{labaar82microw}
Frederik Labaar, \emph{New discriminator boosts phase noise testing},
  Microwaves \textbf{21} (1982), no.~3, 65--69.

\bibitem[LSL84]{lance84}
Algie~L. Lance, Wendell~D. Seal, and Frederik Labaar, \emph{Phase noise and
  {AM} noise measurements in the frequency domain}, Infrared and Millimeter
  Waves (Kenneth~J. Button, ed.), vol.~11, Academic Press, New York, NY, 1984,
  pp.~239--284.

\bibitem[Pap92]{Papoulis:probability}
Athanasios Papoulis, \emph{Probability, random variables and stochastic
  processes}, 3rd ed., Mc{G}raw Hill, New York, 1992.

\bibitem[RB07]{rubiola2007uffc-am-to-pm-pollution}
Enrico Rubiola and Rodolphe Boudot, \emph{The effect of {AM} noise on
  correlation phase noise measurements}, IEEE Trans.\ Ultras.\ Ferroelec.\ and
  Freq.\ Contr. \textbf{54} (2007), no.~5, 926--932.

\bibitem[RG00]{rubiola2000rsi-correlation}
Enrico Rubiola and Vincent Giordano, \emph{Correlation-based phase noise
  measurements}, Rev.\ Sci.\ Instrum. \textbf{71} (2000), no.~8, 3085--3091.

\bibitem[RG02]{rubiola2002rsi-matrix}
\bysame, \emph{Advanced interferometric phase and amplitude noise
  measurements}, Rev.\ Sci.\ Instrum. \textbf{73} (2002), no.~6, 2445--2457,
  Also http://arxiv.org, document ar{X}iv:physics/0503015v1.

\bibitem[RSHM05]{rubiola2005josab-delay-line}
Enrico Rubiola, Ertan Salik, Shouhua Huang, and Lute Maleki, \emph{Photonic
  delay technique for phase noise measurement of microwave oscillators}, J.
  Opt. Soc. Am. B - Opt. Phys. \textbf{22} (2005), no.~5, 987--997.

\bibitem[Rub05]{rubiola2005arxiv-am-noise}
Enrico Rubiola, \emph{The measurement of {AM} noise of oscillators},
  http://arxiv.org, document ar{X}iv:physics/0512082, December 2005.

\bibitem[Rut78]{rutman78pieee}
Jacques Rutman, \emph{Characterization of phase and frequency instabilities in
  precision frequency sources: Fifteen years of progress}, Proc.\ IEEE
  \textbf{66} (1978), no.~9, 1048--1075.

\bibitem[San68]{sann68mtt}
Klaus~H. Sann, \emph{The measurement of near-carrier noise in microwave
  amplifiers}, IEEE Trans.\ Microw.\ Theory Tech. \textbf{9} (1968), 761--766.

\bibitem[SCJ{\etalchar{+}}07]{salzenstein2007appa-dual-delay-line}
Patrice Salzenstein, Johann Cussey, Xavier Jouvenceau, Herv\'{e} Tavernier,
  Laurent Larger, Enrico Rubiola, and G\'{e}rard Sauvage, \emph{Realization of
  a phase noise measurement bench using cross correlation and double optical
  delay line}, Acta Phys.\ Polonica A \textbf{112} (2007), no.~5, 1107--1111.

\bibitem[SFF99]{sampietro99rsi}
M.~Sampietro, L.~Fasoli, and G.~Ferrari, \emph{Spectrum analyzer with noise
  reduction by cross-correlation technique on two channels}, Rev.\ Sci.\
  Instrum. \textbf{70} (1999), no.~5, 2520--2525.

\bibitem[SLSS03]{Spietz2003Science}
Lafe Spietz, K.~W. Lehnert, I.~Siddiqi, and R.~J. Schoelkopf, \emph{Noise of a
  tunnel junction primary electronic thermometry using the shot noise of a
  tunnel junction}, Science \textbf{300} (2003), 1929--1932.

\bibitem[SYMR04]{salik04fcs-xhomodyne}
Ertan Salik, Nan Yu, Lute Maleki, and Enrico Rubiola, \emph{Dual
  photonic-delay-line cross correlation method for the measurement of microwave
  oscillator phase noise}, Proc. Europ.\ Freq.\ Time Forum and Freq.\ Control
  Symp. Joint Meeting (Montreal, Canada), August~23-27 2004, pp.~303--306.

\bibitem[VA89]{vanier:frequency-standards}
Jacques Vanier and Claude Audoin, \emph{The quantum physics of atomic frequency
  standards}, Adam Hilger, Bristol, UK, 1989.

\bibitem[{Vig}99]{ieee99std1139}
John~R. {Vig (chair.)}, \emph{{IEEE} standard definitions of physical
  quantities for fundamental frequency and time metrology--random instabilities
  ({IEEE} standard 1139-1999)}, {IEEE}, New York, 1999.

\bibitem[VMV64]{vessot64nasa}
R.~F.~C. Vessot, R.~F. Mueller, and J.~Vanier, \emph{A cross-correlation
  technique for measuring the short-term properties of stable oscillators},
  Proc.\ {IEEE-NASA} Symposium on Short Term Frequency Stability (Greenbelt,
  {MD}, {USA}), November~23-24 1964, pp.~111--118.

\bibitem[VSHK89]{verbruggen89apa}
A.~H. Verbruggen, H.~Stoll, K.~Heeck, and R.~H. Koch, \emph{A novel technique
  for measuring resistance fluctuations independently of background noise},
  Applied Physics A \textbf{48} (1989), 233--236.

\bibitem[WSGG76]{walls76fcs}
F.~L. Walls, S.~R. Stain, J.~E. Gray, and D.~J. Glaze, \emph{Design
  considerations in state-of-the-art signal processing and phase noise
  measurement systems}, Proc.\ Freq.\ Control Symp. (Atlantic City, {NJ},
  {USA}), June~2-4 1976, pp.~269--274.

\bibitem[YM96]{Yao1996josab-oeo}
X.~Steve Yao and Lute Maleki, \emph{Optoelectronic microwave oscillator}, J.
  Opt. Soc. Am. B - Opt. Phys. \textbf{13} (1996), no.~8, 1725--1735.

\end{thebibliography}

\end{document}